\documentclass[fleqn,usenatbib]{mnras}

\usepackage[T1]{fontenc}
\usepackage{ae,aecompl}

\usepackage{graphicx}	
\usepackage{amsmath}	
\usepackage{amssymb}	
\usepackage{accents}
\usepackage{multirow}
\usepackage{newtxtext,newtxmath}

\title[NSC formation efficiency]{
On the link between nuclear star cluster and globular cluster system mass, nucleation fraction and environment}

\author[Leaman \& van de Ven]{
Ryan Leaman$^{1,2}$\thanks{E-mail: ryan.leaman@univie.ac.at} and
Glenn van de Ven,$^{2}$
\\
$^{1}$Max-Planck Institut f\"ur Astronomie, K\"onigstuhl 17, D-69117 Heidelberg, Germany\\
$^{2}$Department of Astrophysics, University of Vienna, T\"urkenschanzstrasse 17, 1180 Wien, Austria\\
}

\date{Accepted XXX. Received YYY; in original form ZZZ}

\pubyear{2015}

\begin{document}
\label{firstpage}
\pagerange{\pageref{firstpage}--\pageref{lastpage}}
\maketitle

\begin{abstract}
We present a simple model for the host mass dependence of the galaxy nucleation fraction ($f_{nuc}$), the galaxy's nuclear star cluster (NSC) mass and the mass in its surviving globular clusters ($M_{GC,obs}$). 
Considering the mass and orbital evolution of a GC in a galaxy potential, we define a critical mass limit ($M_{GC,lim}$) above which a GC can simultaneously in-spiral to the galaxy centre due to dynamical friction and survive tidal dissolution, to build up the NSC.
The analytic expression for this threshold mass allows us to model the nucleation fraction for populations of galaxies. We find that the slope and curvature of the initial galaxy size-mass relation is the most important factor (with the shape of the GC mass function a secondary effect) setting the fraction of galaxies that are nucleated at a given mass.  The well defined skew-normal $f_{nuc} - M_{gal}$ observations in galaxy cluster populations are naturally reproduced in these models, provided there is an inflection in the {initial} size-mass relation at $M_{gal} \sim 10^{9.5} {\rm M_{\odot}}$.  
  Our analytic model also predicts limits to the $M_{gal} - M_{GC,tot}$ and $M_{gal} - M_{NSC}$ relations which bound the scatter of the observational data.  Moreoever, we illustrate how these scaling relations and $f_{nuc}$ vary if the star cluster formation efficiency, GC mass function, galaxy environment or galaxy size-mass relation are altered. Two key predictions of our model are: 1) galaxies with NSC masses greater than their GC system masses are more compact at fixed stellar mass, and 2) the fraction of nucleated galaxies at fixed galaxy mass is higher in denser environments. That a single model framework can reproduce both the NSC and GC scaling relations provides strong evidence that GC in-spiral is an important mechanism for NSC formation. 
\end{abstract}

\begin{keywords}
galaxies: evolution -- galaxies:star clusters: general -- galaxies: structure
\end{keywords}



\section{Introduction}
Some of the highest baryonic densities in the universe occur in the centres of galaxy gravitational potentials.  Super-massive black holes (SMBHs), hot accretion disks and stellar clusters are all commonly found in these regions, and offer a clue to the formation history of the host galaxy (c.f., \citealt{Magorrian98}).  For example, the growth of the galactic potential, as traced by stellar mass in the bulge or velocity dispersion, is tightly correlated with the black hole mass, despite the latter having a much smaller radius of influence than any galactic scale of interest \citep{Haring04,Remco16,Schutte19}

Of the three central massive objects, nuclear star clusters (NSCs) are the most easily observed and characterized.  They are identified by a photometric light excess in the galaxy's surface brightness profile - on top of the smooth exponential disk and/or Sersic bulge component of the galaxy (e.g., \citealt{Boeker02}).

These stellar agglomerations are found in galaxies of all morphological types with stellar mass $10^{6} \leq M_{*,gal} \leq 10^{12}$.  Their size tends to be correlated with their mass, and the mass ratio of NSC to host galaxy mass decreases monotonically (albeit with large scatter), as galaxy mass increases \citep{Georgiev16}.  Their photometric colours indicate primarily old stellar populations, however most show evidence of metal rich, young populations as well \citep{Boeker04,Kacharov18}

NSCs are typically more massive than globular clusters (GCs), but of comparable size, implying surface mass densities in excess of $\Sigma_{*} \gtrsim 10^{5} {\rm M_{\odot}} {\rm pc}^{-2}$ \citep{Norris14}.  Their possible formation from \textit{in-situ} gas inflows and star formation has been proposed since \citep{Loose82}, and recent observations of young stars in the MW's own NSC have shown evidence for this process\\
\\
\citep{Levin03,Bartko09,Genzel10,FeldmeierKrause15,Paco20}. However their high densities are not simple extensions of the surface mass profiles of the central bulges (which have $\Sigma \lesssim 10^{4} {\rm M_{\odot}} {\rm pc}^{-2}$) in galaxies - and this is one piece of evidence, along with their predominantly old ages, that `external' mechanisms may play a role in their formation.

\subsection{In-spiralling globular clusters}
\cite{Tremaine75} postulated their formation could occur by way of GCs in-spiraling due to dynamical friction and later merging.  Subsequent works have shown the validity of this mechanism \citep{CD93, Miocchi06, Bekki04}, demonstrating its ability to reproduce the structural and mass scaling relations of NSCs \citep{Antonini}, as well as the detailed kinematic properties of the most well studied NSC in our own MW \citep{Perets14,Sassa17}

This formation channel necessitates that the GCs are massive enough that dynamical friction can cause their in-spiral from an initial radius, to the galactic centre, within a Hubble time (e.g., \citealt{ArcaSedda14}).  Alternatively the GCs could be brought in with merging satellite galaxies \citep{Gnedin14}, however it is thought that dynamical friction will cease to operate at a radius where the enclosed mass is proportional to the satellite mass \citep{Read06}.  This may prevent efficient delivery of GCs to the galactic centre via dwarf galaxy orbital decay, and therefore star clusters born in the host galaxy (inner disk) may be most likely to form the population of successful in-falling GCs.

The galactocentric radius where these GCs would form, is not \textit{a priori} known - however it would likely need to be in a regime where there was sufficiently high gas density \citep{Kruijssen15}, and low shear \citep{Meidt13} to allow large giant molecular clouds to form.  While the radial dependence of the conditions necessary for massive star cluster formation have been elegantly presented in \cite{ReinaCampos17}, it is still difficult to have complete knowledge of these dynamical states and baryonic densities at the early epochs when an individual galaxy formed their oldest star clusters.  Suffice to say that both GC formation, and NSC growth via in-spiral of GCs may be aided if some GCs form close ($\leq R_{e}$) to the galactic centre to begin with.  

It is important to note that this process is not restricted to a particular epoch, and migration of dense star clusters may occur over an entire Hubble time, provided the gas densities are high enough for them to form.  As shown by \cite{Guillard16} some of these star clusters may migrate with significant gas reservoirs as well.  Both factors (redshift, gas content) suggest that GCs contributing to the nuclear star cluster need not be exclusively old and metal poor (see also \citealt{OrdenesBriceno18}).  

A parallel evolutionary timescale to consider, relates to the internal structural evolution of a GC sitting in an arbitrary tidal field.  Relaxation driven mass-loss will occur for any GC, and this can be accelerated while it is in a strong (changing) gravitational potential \citep{SpitzerHarm58}.  The mass loss rate depends on the internal properties of the GC (stellar mass function, binary fraction, size), as well as its environment.  

Together these imply that the GC must be both, massive enough to avoid complete evaporation as it moves towards the galactic centre, and massive enough to efficiently in-spiral within a Hubble time due to dynamical friction.

The \textit{relative} size of a galaxy (for example the effective radius, $R_{e}$ enclosing half the mass of the galaxy), and how this changes with galaxy mass, will therefore play an important part in both dynamical friction and evaporative timescales.  For a fixed gas density threshold for massive star cluster formation, the galaxy effective radius may set the initial distance which the GC must in-spiral from \citep{Leung19, Leaman20}.  Similarly, the galaxy's structural properties (e.g., concentration, size-mass relation), will determine the strength of the tidal field driving the GC evaporative mass loss \citep{Lamers05,Renaud11,Webb14,Contenta18, Meadows20}. Given the above theoretical framework for NSC formation, it is therefore expected that the differential variation of galaxy size and mass, will play a significant role in how efficiently a galaxy of a given mass can form an NSC. 

Recently, \citealt{RSJ19} (hereafter RSJ19), reported the fraction of galaxies with an NSC, $f_{nuc}$, as a function of galaxy mass in the Virgo cluster.  Their deep imaging data from the NGVS survey \citep{Ferrarese12} along with literature studies of the galaxy nucleation fraction in three other clusters, showed that $f_{nuc}$ follows a well defined skew-normal shape when plotted versus the log of host galaxy stellar mass.  The peaked $M_{gal}-f_{nuc}$ relation suggests a strong galaxy mass-dependent efficiency for the formation and/or survival of NSCs.  The authours in that work speculated that the inefficiency of NSC formation at the low mass end was due to the lack of high mass GCs being formed in low mass dwarf galaxies, while at the high mass end, SMBHs may inhibit NSC growth.  Comparison in that paper to theoretical models invoking GC in-spiral, or NSC formation from in-situ gas both struggled to reproduce the functional form of this relation.

\subsection{Model framework}
Given the intriguing mass-dependent nucleation fraction observed in a variety of environments \citep{Neumayer20}, in this work we present a model for galaxy NSC formation efficiency with an intuitive link to the host galaxy structural and star cluster properties.  Predicting both the GC and NSC populations in a galaxy, as in the seminal work of \cite{CD93}, provides strong constraints on many evolutionary processes for star clusters and their hosts.  In this model, rather than tackle a comprehensive set of physical processes for the buildup of nuclei as in more advanced and computationally expensive numerical studies \citep{Antonini,Guillard16}, we consider simplified expressions for GC mass loss in an evolving tidal field to see how far this single process can reproduce a variety of star cluster - host galaxy scaling relations before it requires extra ingredients.

Our scope and requirements for this model framework are the following:
\begin{itemize}
\item  Self-consistently predict the nucleation fraction \textit{and} NSC and GC system masses of galaxies of all masses
\item Allow for a variation in these due to host environment and structure
\item Predict the scatter and limits for these scaling relations, including uncertainties due to the GC formation properties
\end{itemize}

Subsequent papers in this series will extend the models to incorporate predictions for the fraction of NSC mass that may be formed due to in-situ star formation, predict BH masses, and predict stellar population properties for the NSCs. The above listed aspects of the current model are outlined in this paper in Section 3, while the comparison to data and impact of environment and star cluster properties are discussed in Sections 4 and 5.

\section{Data}
Observational data for galaxy sizes (effective radii, $R_{e}$) are taken from the compilation of \cite{Norris14} which focused on dispersion supported systems in the nearby universe.  We add to this galaxies from the updated Local Volume catalogue of \cite{Alan12}, predominantly low mass galaxies in the Fornax cluster from \cite{Munoz15} and in the Coma cluster from \cite{denBrok14}.  We complement this with published sizes and masses of ultra-diffuse galaxies (UDGs) from \cite{vandokkum14,Yagi16,Roman17a,Roman17b,Lim18,Eigenthaler18}.  We have adopted the distance to the cluster environments as published in the original studies. In cases where only photometric magnitudes or luminosities are recorded we have converted these to stellar masses assuming a stellar $M/L = 2$ for consistency and simplicity (c.f., \citealt{Alan12}).

The binned galaxy nucleation fraction, NSC masses and stellar masses for galaxies in the Coma, Virgo, Fornax clusters, as well as the MW, M31 and M81, are taken from the compilation of RSJ19 and references therein.  During the proofs stage of this manuscript, several other compilations of Local Volume nucleation fraction data were published in the studies of \cite{Carlsten21,Hoyer21,Zanatta21,Poulain21}, to which we refer interested readers.  

We utilize galaxy masses and nuclear star clusters masses from RSJ19 as well as \cite{Peng06,denBrok14,Baldassare14,Georgiev16,Lim18}.  The NSC masses in \cite{Peng06} are computed following the authours' calibration as a function of $g$ and $i$ band magnitudes and colours.  In all other cases a stellar $M/L = 2$ was adopted.

Galaxy total GC system masses ($M_{GC,tot}$) are taken from \cite{Peng06,Spitler09,Harris13,Baldassare14,denBrok14,Forbes18,Fahrion20,Leaman20,Fahrion20b} and RSJ19.  Where not reported directly as stellar masses, the GC system luminosities are transformed to GC total masses assuming $M/L = 1.6$.  For all other samples the total mass in GCs are computed from the reported GC numbers ($N_{GC}$) assuming an average mass of $M_{GC} = 4\times10^{5} {\rm M_{\odot}}$.  We have only included galaxies from all studies that have relative errors in the number of observed GCs of $\sigma_{N_{GC}}/N_{GC} \leq 5$.

\cite{Amorisco18}, \cite{Lim18} and \cite{Prole19} provided estimates of the number of GCs for ultra-diffuse (UDGs) and/or low surface brightness (LSB) galaxies in the Coma and Fornax clusters, and in the case of \cite{Lim18} the authours present nucleation probabilities for the UDGs in their sample. While in the \cite{Amorisco18} and \cite{Prole19} studies the UDG/LSBs are not visually nucleated, we can ask what a \textit{limit} on their NSC mass would be \textit{if one} of the detected GCs in each galaxy happened to be a nuclear star cluster.   This might be the case in such extended low mass systems where NSCs are known to be significantly off-centre \citep{Georgiev16}, and where the potentially cored DM profiles of the UDGs \cite{diCintio17}  may result in core-stalling halting the dynamical friction driven decay of in-spiraling GCs well outside the gravitational centre \citep{Petts15}. 

To estimate the limiting NSC and GC masses for these UDG/LDG samples we randomly select one of the GCs in each galaxy and assign it a mass from a log-normal GC mass function with mean and width ($5.2, 0.7$).  We then compute the mass of this as the NSC, and the GC system mass as the summation of the remaining cluster population (also sampled from this GC mass function).  This is repeated $10^{4}$ times per galaxy.  In what follows we take the $1-\sigma$ distributions of the values from this stochastic exercise as limits for the NSC and GC system masses of the UDG/LSBs in those studies.  

We note that studies differ on the definition of UDGs (based on size cuts, surface brightness cuts, and whether that surface brightness definition is evaluated in the centre or at the effective radius).  Ongoing efforts in the community to robustly assess the nucleation fraction in complete and unbiased samples of UDGs will be most useful in understanding the evolution of these systems and their nuclei. However for our purposes the studies here provide a sample of extended, low density objects with which to study the limiting effect of galaxy structure on NSC formation efficiency within our model.

In all cases, if not provided by the authours we assume a 0.3 dex uncertainty on $M_{GC,tot}$, $M_{NSC}$ and $M_{gal}$.  This neglects statistical uncertainties in the GC system mass which may come from un-quantified contamination of intra-cluster GCs, or completeness corrections.

\section{Model description}
In what follows, we will model the host galaxy dependencies for two relevant timescales: the timescale for a GCs orbital radius to decay to the galactic centre via dynamical friction, and the timescale for tidal dissolution of a GC.

\subsection{Limiting Masses for In-spiralling GCs}
Analytic dynamical friction estimates approximate the orbital decay of an object (here a GC) while it moves through a sea of slower moving stars, as the system tends towards energy equipartition (c.f., \citealt{Chandrasekhar43}).  The orbital angular momentum of the GC will evolve due to dynamical friction in this approximation as:
\begin{equation}
\frac{\partial L}{\partial t} = R\frac{\partial V}{\partial t} = R\frac{F_{DF}}{M_{GC}}.
\end{equation}

If the GC orbital velocity varies slowly during infall (for example, due to an isothermal density profile), then the differential evolution can be approximated as:
\begin{equation}
R\frac{dR}{dt} = \frac{-GM_{GC}{\rm log}\Lambda f(<V)}{V_{GC}}.
\end{equation}
Here $f(<V)$ is the distribution of stars with velocities less than the GC velocity (which in this approximation is typically the circular velocity of the host galaxy at some radius).  The velocity distribution is most generally taken to be Maxwellian with a scaled velocity of $\nu = V_{c}/\sqrt{2}\sigma$ (though see e.g., \citealt{Leung19} for an example with self-consistent ergodic distribution functions).

We follow \cite{Petts15} in order to model several important dependencies on the host galaxy structure - in particular the functional form of the Coulomb logarithm in a way that is less dependent on the GC mass:
\begin{equation}
{\rm log} (\Lambda) = {\rm log}\left(\frac{{\min}(\rho(R)/\Delta\rho(R), R)}{{\rm max}(r_{GC}, GM_{GC}/V_{GC}^{2}}\right).
\end{equation}

The next timescale to consider, the evaporative mass loss of a GC in a tidal field, is generally described (e.g., \citealt{Spitzer40}) as:
\begin{equation}
\xi(R,t) dt = \frac{-t_{rh}dM}{M_{GC}},
\end{equation}
where $\xi$ is the mass evaporation rate, which has a time dependence due to the internal evolution of the GC stellar populations and structure, and a galactocentric distance dependence due to the external tidal field felt by the GC.  

The internal and external effects will work together to result in a time-evolving mass loss rate, which makes it not possible to simultaneously describe the GC mass and orbital co-evolution with purely analytic descriptions.  While sometimes for simplicity a constant evaporation rate has been adopted, recently \cite{Madrid17} used N-body and Monte Carlo simulations to study the evolution of a star cluster in a strong tidal field, and provided one of the first fitting functions for $\xi(R,t)$ at different distances and times during a cluster's infall in a galaxy potential.  

This calibration of the simultaneous time and distance dependence of $\xi$ is a crucial ingredient which could allow a coarse analytic model for the joint survival and in-spiral of a GC in a mass dependent galactic potential.  While most importantly it let those authours explore the variation in $\xi$ in different locations and phases within a GCs evolution in a galaxy, even for constant $\xi$ we can conceptually use it to guide our estimates of the GC mass loss and how this relates to its orbital evolution at different snapshots in time.  Expressing the GC mass in Equation 2 in terms of the galactic structural parameters gives:
\begin{equation}
M_{GC} = -R\frac{dR}{dt}\frac{V_{GC}}{G{\rm log}\Lambda f(<V)}
\end{equation}
which allows us to equate the mass loss evolution to the orbital decay:
\begin{equation}
\int_{R=0}^{R_{i}}\xi(R|t)R dR = \int_{M(R=0)}^{M(R_{i})}\frac{G{\rm log}(\Lambda)f(<V)t_{rh}}{V_{max}}dM_{GC} \equiv \gamma\Delta M_{GC}
\end{equation}
where $\gamma = G{\rm log}(\Lambda)f(<V)t_{rh}V_{max}^{-1}$ are terms that change weakly with host distance or GC mass \textit{relative} to the change in $\xi$ and hence $M_{GC}$.  For example for fixed GC size, $t_{rh} \propto M_{GC}^{0.4}$, while the other terms are  independent of $M_{GC}$.

As $\xi$ formally evolves with both $R$ and $t$, analytically we can only use the fitting function from \cite{Madrid17} as a guide for possible initial (constant) values of $\xi$. Taking a constant value of $\xi$ as is often adopted is certainly viable if we wish to consider the relatively moderate early evolution in $\xi$ in regions of the galaxy where the baryonic density is not changing drastically. In this case we generally would have an expression for the change in GC mass ($\Delta M = M_{i}-M_{f}$) as it in-spirals to the galaxy centre of the form $\Delta M \sim \xi_{0}R_{i}^{2}\gamma^{-1}$.

This generic form of Equation 6 will be used here to allow for a simple view of how other parameters affect our model predictions for the nuclear star cluster properties.  For a given initial galaxy effective radius ($R_{i}$), a constant value of $\xi_{0}$ can be estimated at fixed time through the fitting function, and integrating the left and hand side of Equation 6 to $R_{0}$ provides an equivalence to the GC mass loss over this cluster infall.  Even though we can't analytically solve this with an evolving $\xi(t)$\footnote{As discussed in \cite{Madrid17}, the variation of $\xi$ in the very final phases of cluster in-spiral can be of order a factor of 30, with the result that as much as $100\%$ of the final mass is lost in an integrated sense - in line with \cite{Webb14}.}, adopting a constant mass-loss as in other studies, lets us examine aspects of the host galaxy dependence on star cluster populations.  This may be especially appropriate for massive clusters where mass loss may not appreciably alter the orbital evolution. 


For such scenarios of nearly constant or negligible mass-loss, this expression effectively removes an explicit orbital time dependence and equates the galactocentric distance evolution directly to the mass loss of the GC, $\Delta M_{GC}$ and distance independent (for the simplest assumptions) properties of the host potential, $\gamma$.  With this and the functional form of $\xi(R,t)$ used to estimate $\xi_{0}$ from the N-body simulations of \cite{Madrid17}, we can then integrate the left hand side of Equation 6 from some initial GC formation distance, $R_{i}$ to the galaxy centre, \textit{to understand the approximate proportional limiting change in a GC's mass due to both internal and tidal effects during infall}.  
The fitting function derived in \cite{Madrid17} was based on $N-$body simulations in MW like potentials.  Future numerical work studying the mass loss rate in a variety of potentials will permit improvements to this analytic exercise, but for now we study the model predictions using the existing formulae, as we still are primarily concerned with the host properties rather than the exact value in our approximation of $\xi_{0} = \xi(R_{i},t_{form})$.  The evaluation of the integral in Equation 6 is shown in full in Appendix \ref{app:a}.  We note that for coefficients in the fit, our model easily reduces to a simpler one tracking orbital in-spiral with no GC mass-loss for examination of some illustrative scaling relations.

Evaluation of Equation 6 forms the backbone to our analytic model, predicting the necessary mass to simultaneously survive tidal effects and successfully migrate to the centre of a host galaxy.  The trade-off in being able to analytically approximate for GCs surviving both processes (dissolution, in-spiral) is that we cannot differentiate between the two nor uniquely associate a single cause for when GCs do not make it to the nucleus.  Here it is \textit{either} due to tidal dissolution, or ineffective migration. Fully separating those processes requires numerical models, but despite this we will focus on what can be learned by considering the joint evolutionary process analytically. 

We finally would like to make a remark on the nomenclature of `GC' in this paper.  Throughout this work we adopt this term for convenience, but the model is agnostic to the type of star cluster which is undergoing migration.  These need not be old, metal poor ``classical'' GCs, but are most likely massive inner disk clusters (e.g., \citealt{Khoperskov18}), potentially even with gas reservoirs \citep{Guillard16}; although see \citealt{Bastian18} for discussion on observational constraints to this scenario).  We only specify the mass for this process to occur and the star clusters to survive.  Studies looking at the age and metallicity properties of NSCs will be crucial in assessing the formation pathways, as well as the properties of any successful migrating clusters \citep{Fahrion21}.

\begin{figure}
\includegraphics[width=0.48\textwidth]{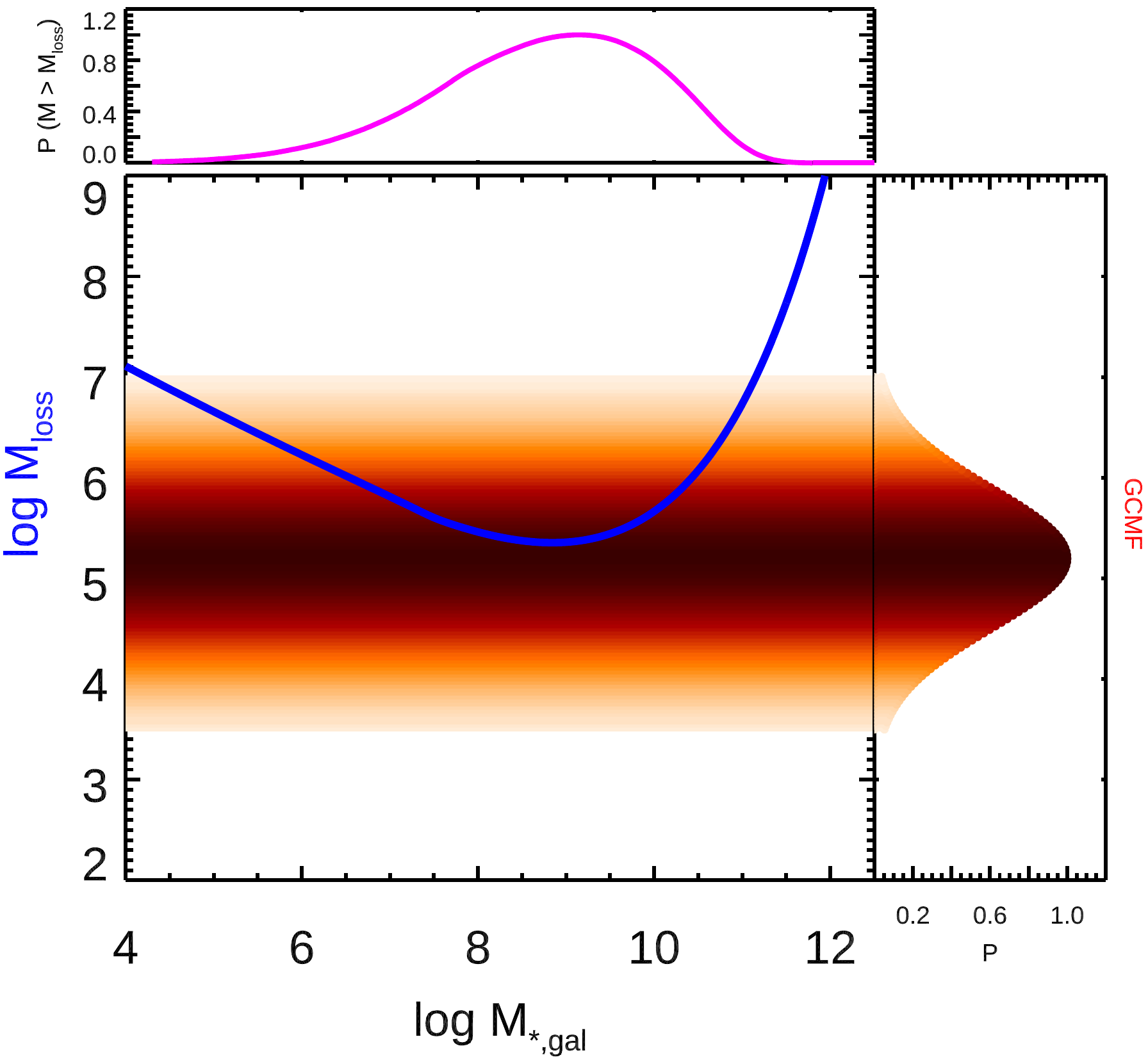}
\caption{Limiting GC mass needed to simultaneously survive mass loss and in-spiral to the centre due to dynamical friction in our model (\textit{blue line}).  Canonical GC log-normal mass function is shown as the shaded density map and histogram in the right panel.  The fraction of the GC mass function above the threshold mass directly maps to nucleation fraction probability (\textit{top panel}).}
\label{fig:f1}
\end{figure}

\subsection{Nucleation Probability}
Equation 6 provides a possible value for the combined mass loss over the lifetime of a GC as it in-spirals, and thus defines a characteristic mass ($\Delta M$) that a GC must at minimum have in order to survive in-spiral.  We should therefore expect GCs with masses at or above this mass limit to contribute to the formation of the NSC.

A dependence on host galaxy structure for $\Delta M$ enters Equation 6 via the starting galactocentric radius of the GC, $R_{i}$.  As discussed in the introduction, there exist regions of a galaxy where its gas surface density and dynamics are conducive to forming massive giant molecular clouds \citep{ReinaCampos17}.  Both will have host mass dependencies, which we take into account by expressing the initial GC galactocentric radius in terms of the galaxy effective radius, $R_{i} = \zeta R_{e}$.  

In this setup the relative scale radius of galaxies of different masses will set the duration and strength of mass loss the GC experiences - and therefore whether it can contribute to an NSC build-up.  While there is variation within a galaxy and from galaxy-to-galaxy in the expected initial distance a GC forms at, there are hydrodynamical arguments why a threshold pressure for dense star cluster formation will occur within a similar fractional scaled radius in galaxies of different masses \citep{Leaman20}, as we adopt here.  We explicitly show the impact of relaxing this assumption in Appendix \ref{app:c}.

Figure \ref{fig:f1} shows the change in initial GC mass ($M_{loss} \equiv \Delta M$) in order for a GC to successfully in-spiral to the nucleus, as a function of host galaxy mass.  This starts each GC at $R_{i} = R_{e}$ in Equation 6, and adopts a baseline galaxy $R_{e}-M_{gal}$ relation derived from \cite{Graham06}.  The most extreme mass thresholds are for GCs in low mass and high mass galaxies, with a minimum around $M_{gal} \sim 10^{9}$

The nucleation probability in our model is computed at each galaxy mass, by asking what fraction of GCs, formed in a canonical log-normal or power-law mass function (e.g., \citealt{Harris91,Gieles09}) are above the joint survival mass threshold $M_{loss}$.  Specifically, $f_{nuc} = N(M_{GC} > M_{loss})/N_{GC,tot}$.  This is shown for an illustrative example as the magenta line in Figure 1, with an example log-normal GC mass function shaded for context at each galaxy mass. In Section 4.2 we show how this illustrative example changes in detail as we relax assumptions on the GC or host galaxy properties.

\subsection{NSC and GC System Masses}
\begin{figure}
\includegraphics[width=0.48\textwidth]{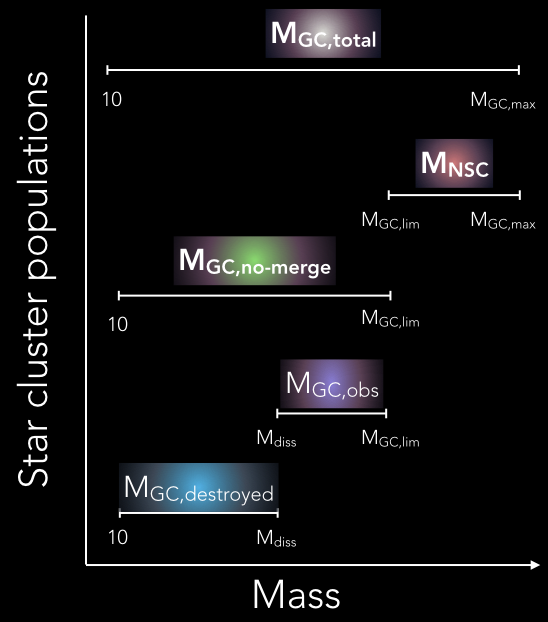}
\caption{Division of a galaxy's star cluster populations into different sub-population masses (\textit{coloured labels}; units of M$_{\odot}$) and limits on the mass function (\textit{bar labels}).  The limiting mass for GC in-spiral and survival predicted in our model $M_{GC,lim}$ sets the lower limit to the integrated NSC mass, and also the upper limit to the most massive observed present day GC outside the nucleus.}
\label{fig:nscflow}
\end{figure}

A goal for our model is to use the same limiting mass  which dictates the probability of nucleation, to predict the NSC mass and total mass in observed GCs within a galaxy.  To begin we subdivide the total star cluster populations formed \textit{in-situ} in a galaxy, into various categories, as shown in Figure \ref{fig:nscflow}.

The galaxy's total mass in GCs ever formed \textit{in-situ}, $M_{GC,tot}$, is comprised of star clusters of masses from some lower limit ($M_{cl,min} = 10$ M$_{\odot}$ as a default, though the results are insensitive to the precise value) up to the most massive star cluster ever formed, $M_{cl,max}$.  From dynamical friction arguments the most massive GC observed at present day should not be more massive than the NSC, but we leave this quantity flexible for the moment. Following \cite{Elmegreen18}, for a power law distribution of slope $\alpha = -2$ we can write the expected total mass in star clusters formed as:
\begin{equation}
    M_{GC,total} = M_{cl,max}\left(1 + {\rm ln}\left(\frac{M_{cl,max}}{M_{cl,min}}\right)\right)
\end{equation}
This total mass in clusters is some fraction $\eta$ of the mass the galaxy ever formed in stars, such that the total mass in clusters can be expressed as:
\begin{equation}
    M_{GC,total} = \eta M_{*,gal}
\end{equation}

Our model predicts that only star clusters \textit{above} a limiting mass $\Delta M$ are able to simultaneously in-spiral and survive to build up an NSC.  However Equation 6 is the limiting case where the cluster makes it to the centre with zero mass remaining.  In practice we are interested in the cases where $M_{f} = M(R=0) > 0$, and so the condition for survival and in-spiral will be some factor $F$ larger than $\Delta M$.  Chemical \citep{Larsen12} and dynamical \citep{Web15} studies suggest that most GCs should not have been significantly ($F \lesssim 3.5$) times more massive at birth otherwise runaway dissolution, un-observed mass functions and contributions of chemically anomalous populations to field stars in galaxies would be inconsistent with current observational constraints.  Hence we consider that the actual limiting star cluster mass from Equation 6 will be $M_{GC,lim} = 3.5\Delta M$.  However we note that this does not affect the proportionality with host properties in Equation 6.

The \textit{most} massive GC that could in-spiral will be equal to the the most massive one ever formed, $M_{cl,max}$ which enters in Equation 7. The expectation value for the NSC mass is then:
\begin{equation}
M_{NSC} =  M_{cl,max}\left(1 + {\rm ln}\left(\frac{M_{cl,max}}{M_{GC,lim}}\right)\right)
\end{equation}

In the case where the galaxy is nucleated, the maximum cluster mass is guaranteed to be above the threshold mass for in-spiral, $M_{cl,max} \geq M_{GC,lim}$.  For these galaxies with both NSCs and GCs we can then combine Equations 7-9 to rewrite the expected NSC mass as:
\begin{equation}
    M_{NSC} = \eta M_{*,gal}\left[\frac{1 + {\rm ln}\left(\frac{M_{cl,max}}{M_{GC,lim}}\right)}{1 + {\rm ln}\left(\frac{M_{cl,max}}{M_{cl,min}}\right)}\right]
\end{equation}

The remainder of the GCs which are below the threshold mass for in-spiral, make up the field population of star clusters in the galaxy and range in mass from $M_{cl,min}$ to $M_{GC,lim}$.  Not all of these survive until present day, with those below a mass $M_{diss}$ dissolving due to evaporative relaxation effects of the GC in its long-term (predominantly low density) environment, or due to tidal threshing in the dense giant molecular cloud complexes where it may be born.  The shorter of these two timescales is what is relevant here, however both are expected to vary over time and location within a galaxy \citep{Kruijssen15}.  The surviving population of GCs which make up the galaxy's GC system observed at present is:

\begin{equation}
    M_{GC,obs} = \eta M_{*,gal} - M_{NSC} - M_{diss}\left(1 + {\rm ln}\left(
    \frac{M_{diss}}{M_{cl,min}}\right)\right)
\end{equation}

The dissolution mass is generally difficult to know, and the biased population of surviving star clusters which make it to the NSC may be efficient at migrating out of dense molecular gas layers and less subject to the cluster mortality that the general population undergoes in high gas fraction environment.  However Equation 10 shows that for our model this is not an explicit factor for predicting the NSC mass.  While the dependence of the dissolution mass on the final GC system mass is non-negligible for individual systems, with a 1.5 dex increase in $M_{diss}$ resulting in a reduction of $0.2-0.7$ dex in $M_{GC,obs}$ for an LMC type galaxy - it would appear this variation does not constitute the full scatter of the scaling relation presented in Figure \ref{fig:obsopt}

\begin{figure*}
\includegraphics[width=0.98\textwidth]{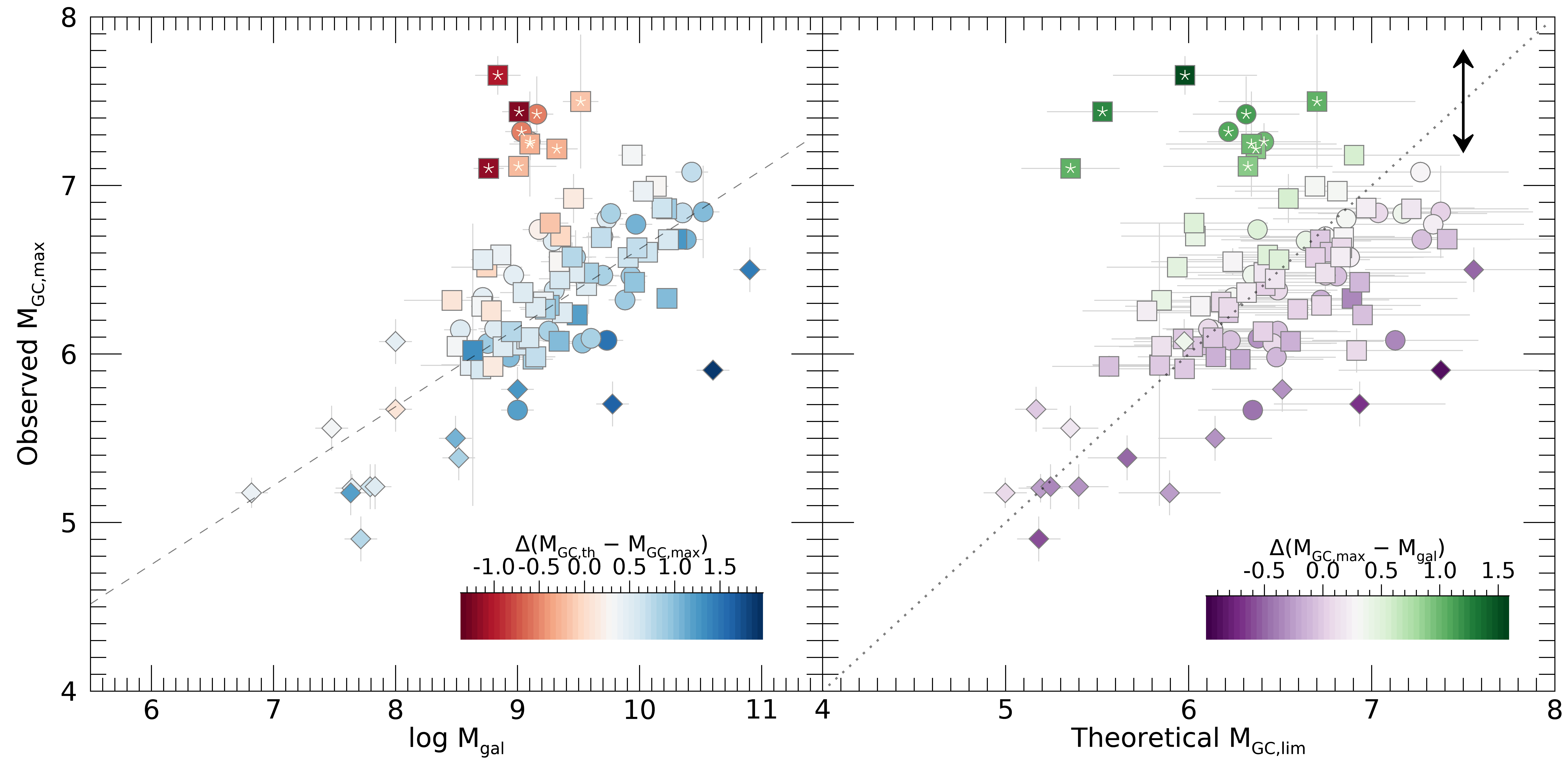}
\caption{\textit{Left:} Most massive GC observed in each galaxy for galaxies in the Local Group (\textit{diamonds}), Fornax (\textit{circles}) and Virgo cluster (\textit{squares}) as a function of host galaxy stellar mass. \textit{Right:} Predicted limiting GC mass beyond which GCs in-spiral to the NSC, versus the most massive observed GC in each galaxy.  The colour coding in each panel shows the residuals from the opposing panel.  Seven objects which are clear outliers at fixed galaxy mass are also outliers in the right panel (\textit{asterisks}) and are likely stripped nuclei rather than GCs.  Vertical arrow shows systematic uncertainty due to distance variations within the galaxy clusters.}
\label{fig:modval}
\end{figure*}

Equations 9 and 11 present the expectation values for the NSC and GC system masses, however we can also consider some limiting cases.  Both the NSC mass and the GC total mass should not be larger than the galaxy mass times the bound cluster fraction $\eta$ (Equation 8).  The lowest NSC mass possible would correspond to a single GC with mass $M_{GC,lim}$ in-spiralling to the galaxy centre to form a nucleus.  The minimum total mass in GCs for a galaxy with a nucleus, will have a power law distribution with common limits at $M_{diss}$ and $M_{GC,lim}$. Together these bound the range of total mass in observed GCs and the NSC mass expected in a galaxy:
\begin{align}
M_{GC,lim} &\leq M_{NSC} \leq \eta M_{*,gal}\\
M_{GC,lim}\left(1 + {\rm ln}\left(\frac{M_{GC,lim}}{M_{diss}}\right)\right) &\leq M_{GC,obs} \leq \eta M_{*,gal}
\end{align}

Together equations 10-13 can be used with our model prediction for $M_{GC,lim}$ to describe the expected and limiting masses in observed GCs and NSCs in galaxies of an arbitrary mass.  The input parameters are the galaxy size-mass relation which enters into the computation for the limiting GC mass for in-spiral ($M_{GC,lim}$) in our model.  The free parameters are the limiting GC mass which survives tidal dissolution, the most massive GC ever formed, and the fraction of the galaxy's star formation in star clusters: $M_{diss}$, $M_{cl,max}$ and $\eta$ respectively.

\subsection{Model Summary}
The simple model here considers one possible pathway for the build-up of NSCs - namely the in-spiral of massive star clusters via dynamical friction.  Our model is by no means exhaustive, and neglects well known effects from in-situ star formation and destruction of NSCs by SMBH binaries.  However our goal is to explore some key parameters which are relevant for this pathway, and the consequences their expected variation with host galaxy mass have on the \textit{joint} NSC, GC system mass, and nucleation fraction probability of populations of galaxies.

The observable inputs used are the host galaxy stellar mass $M_{*,gal}$, and its effective radius $R_{e,gal}$.  The free parameters which can be varied by the user are:  average initial formation distance for in-spiraling GCs ($R_{i}/R_{e,gal}$), fraction of the galaxy's mass ever formed in star clusters ($\eta$), an estimate of the typical mass cluster which survives dissolution processes ($M_{diss}$), and parameters describing either a log-normal ($< M_{GCMF} >$, $\sigma_{GCMF}$) or power law star cluster mass function ($M_{cl,min}$, $M_{cl,max}$, $\alpha$).  The model then predicts: the limiting mass above which GCs in-spiral and contribute to the NSC growth ($M_{GC,lim}$), the probability of the galaxy being nucleated ($f_{nuc}$), and limits and expectation values for the final mass in the NSC due to in-spiralling GCs ($M_{NSC}$) and the mass in surviving star clusters ($M_{GC,obs}$).

\section{Results}
In this section we show the comparison of our model to the observational data for galaxy nucleation fraction and the GC and NSC mass scaling relations.  In the appendices we quantify the dependence of the model predictions on galaxy structural relations and cluster formation efficiency.

\subsection{Limiting GC masses for in-spiral}
Figure \ref{fig:modval} shows a comparison between the \textit{observed} most massive surviving GC, and the limiting GC mass to contribute to NSC formation in our model $M_{GC,lim}$, for a sample of Local Volume, Virgo and Fornax cluster galaxies (see Fahrion et al., in prep. and references there within).  The theoretical GC limiting mass is predicted only from the host galaxy properties (size, mass) and shows excellent agreement with the mass of the most massive observed GCs in these galaxies.  Seven objects which are outliers in this relation, also are seen to be outliers in the $M_{GC,max} - M_{gal}$ relation, and are likely stripped nuclei rather than massive GCs. Excluding these, we find zero offset between the predicted and empirical limiting GC masses, with an intrinsic scatter of 0.26 dex abou the 1:1 relation. Figure \ref{fig:modval} provides a crucial validation of a key quantity underpinning our model - namely the dividing mass $M_{GC,lim}$ beyond which GCs contribute to NSC growth.

\subsection{Galaxy nucleation fraction}
\begin{figure}
\includegraphics[width=0.48\textwidth]{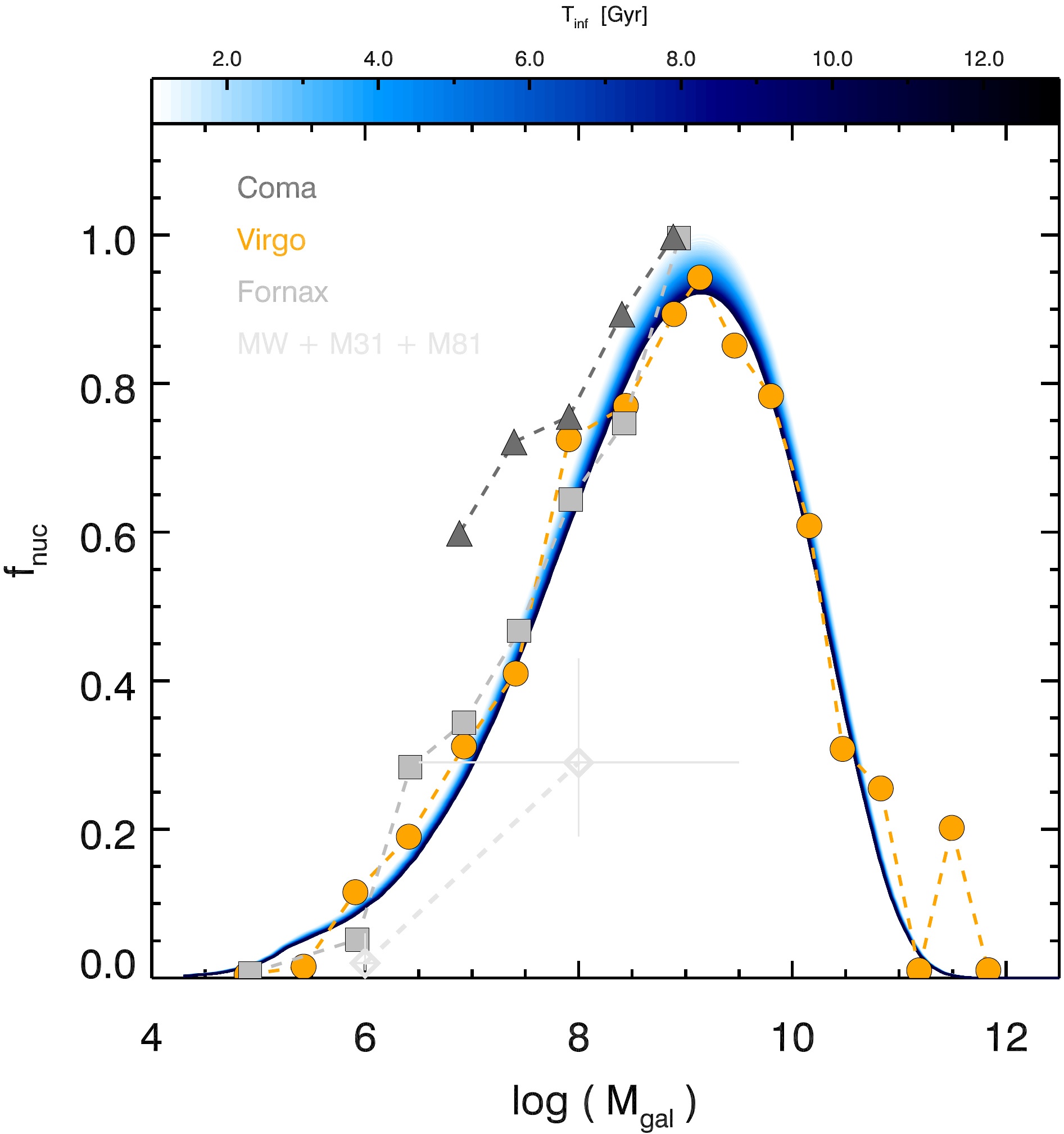}
\caption{Nucleation fraction as a function of host galaxy mass.  Data points are from \protect{\citealt{RSJ19}} for galaxies in four clusters/groups.  Our baseline model is shown as the coloured curves, for different starting times for the GC infall process.}
\label{fig:f2}
\end{figure}

Figure \ref{fig:f2} shows the nucleation fraction as a function of galaxy mass for our baseline models.  These use the $R_{e}-M_{gal}$ relation from \cite{Graham06}, and take the initial GC galactocentric radius to be $R_{i} = R_{e}$.  Equation A2 allows for a time offset in evaporation rate, for example if the GC formed at a younger relative age.  This parameter is left free in Figure \ref{fig:f1} to show the minimal impact that GC age produces, as the destruction and in-spiral timescales in unsuccessful cases are quite robust for the case of $R_{i} = R_{e}$ (e.g., cluster disruption will not be the dominant factor in these cases).  The baseline model shows a remarkable agreement to the skew and peak location of the observed $f_{nuc}$ trends compiled in RSJ19, especially considering it is not a fit to the data but takes as input only properties of the galaxy size.

What factors are driving this simple model to reproduce the location of peak nucleation fraction?  Below and in Appendix \ref{app:c} we discuss which factors may be important in the context of our model framework.

\subsubsection{Dependence on the GC mass function}
Our nucleation fraction is formulated by asking what fraction of GCs form above a host dependent survival criteria.  Thus, we expect a degenerate behaviour between the galaxy dependent survival mass $M_{GC,lim}$, and the mass function that a population of GCs are formed with.

\begin{figure*}
\includegraphics[width=0.98\textwidth]{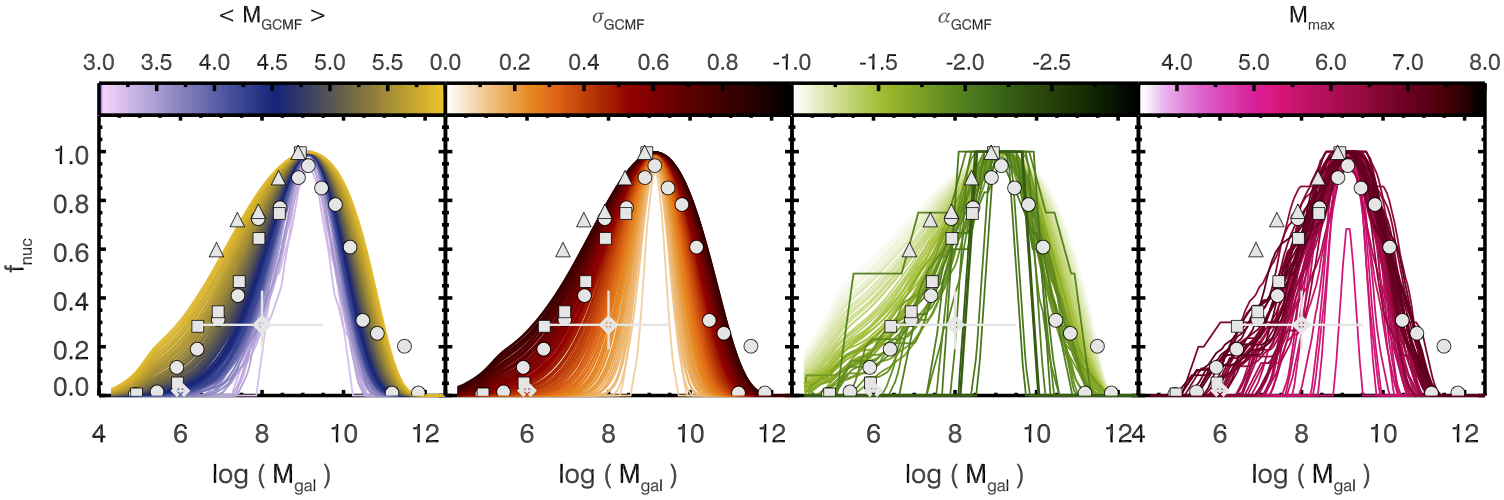}
\caption{Impact of the GC mass function on the galaxy nucleation fraction. Models show the change in $f_{nuc}$ if the GC mass function parameters are varied (from left to right): average mass of log-normal mass function, standard deviation of log-normal mass function, slope of power-law mass function, maximum mass cut-off of power-law mass function. Observations are shown in grey as in Figure \ref{fig:f2}.}
\label{fig:f4}
\end{figure*}

In Figure \ref{fig:f4} we show the dependence of the nucleation fraction predictions on the form of the initial GC mass function.  The left two panels show that the nucleation fraction in galaxies becomes higher if a log-normal GC mass function is shifted to higher average mass, or has a broader distribution.  Similarly, a power law mass function will produce higher nucleation fractions in the galaxies when it becomes shallower, or has a higher maximum mass cutoff. 

These results are both intuitive given how our model is constructed, and can be understood in the context of Figure \ref{fig:f1}.  A smaller characteristic GC mass will shift the log-normal distribution to lower values relative to the threshold mass ($M_{GC,lim}$ ; \textit{blue line}).  This results in a smaller fraction of GCs in a log-normal mass function which are born with masses above this survival threshold.
The functional form of the GC mass function provides a second order change in the width of the $f_{nuc} - M_{gal}$ relation, but does not strongly impact its skewness or peak location.

\subsubsection{Dependence on host galaxy structure}
In addition to the properties of the GC mass function, the mass dependence of galaxy structural properties (density profile shape, effective radius) should strongly effect the in-spiral time and the tidal environment for the GC.  Therefore we expect a strong dependence on the nucleation fraction with galaxy structural scaling relations in our simple model.

Our toy model in Figure \ref{fig:f1} used an example size-mass relation similar to \cite{Graham06}, which has an inflection point in the $R_{e} - M_{gal}$ curve (if extrapolated beyond their lowest mass observations). At face value the inflection in this toy model occurs at approximately the same location as the $f_{nuc}-M_{gal}$ peak.  This is suggestive that the shape of the latter may be driven by the form of the galaxy size-mass relation.

\begin{figure*}
\includegraphics[width=0.98\textwidth]{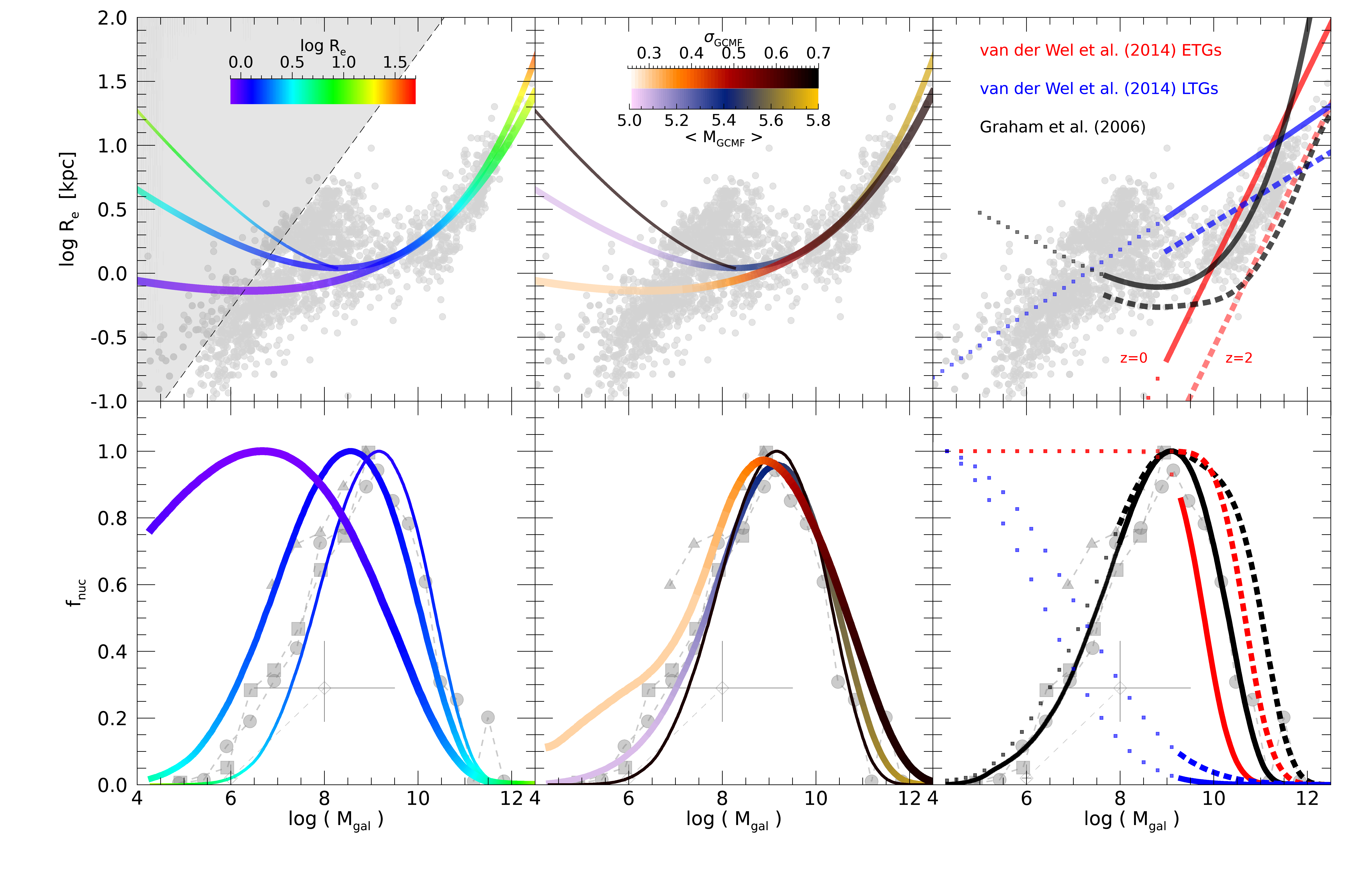}
\caption{\textit{Left:} Response of the nucleation fraction to variations in size-mass relations of the host galaxies. Long dashed line corresponds to a surface brightness limit of $27 {\rm mag ~arcsec}^{-2}$, with a compilation of galaxy sizes indicated as grey points (\textit{see text for references}). Observational data for the nucleation fractions are shown in grey following Figure \ref{fig:f2}. Three illustrative variations are shown in the top row, with the corresponding predicted model nucleation fractions in the bottom row. \textit{Middle:} Degenerate reproduction of the nucleation fraction data are possible for the same input size-mass relations if the parameters of the GC mass function are allowed to vary with host mass. \textit{Right:} Two literature size-mass compilations are shown for reference. Dashed lines show the size-mass relations nucleation fractions if the NSC formation preceded the dominant epoch of size growth and occurred at $z \sim 2$.  }
\label{fig:f5}
\end{figure*}

In the left panel Figure \ref{fig:f5} we quantify this by showing the model nucleation fraction if we adopt three arbitrary galaxy size-mass relations which vary in their low mass slope where observational constraints are difficult.  A heterogeneous sample of galaxy observed sizes (see Section \S 2) are shown as the grey dots, and together with the three example relations, illustrate that a variety of plausible behaviours of the size-mass variation may be allowed at the low mass end.  The corresponding predicted nucleation fractions are shown in the bottom panel. The observational size-mass data in the left panel of Figure \ref{fig:f5} is approximately bisected by the intermediate relation, at least until a surface brightness limit of $\mu = 27~ {\rm mag ~arcsec}^{-2}$ - suggesting that our fortuitous first choice of a galaxy size-mass relation may be a reasonable description for an ensemble population of cluster and field galaxies.

However, while a shallower size-mass relation would seem to over-predict the nucleation probability at low masses ($M_{*,gal} \lesssim 10^{8}$), the models in the left panels are assuming the same GC mass function parameters for galaxies of all masses $(< M_{GCMF} >$,$\sigma_{GCMF} = 5.2, 0.6$).  The middle panel of Figure \ref{fig:f5} explicitly shows how for the \textit{same} size-mass relations, the nucleation fraction may be reduced in the dwarf regime if there are host-mass dependent variations in the GC mass function parameters as seen in e.g., \citealt{Jordan07} where the spread and average GC masses both decrease with host mass.

We are not advocating for any particular choice here - the exact form of the size-mass relation and mass dependence of the GCMF reported across studies has large variance.  We simply wish to show in this figure that there are plausible combinations of both which can go some distance in reproducing the nucleation fraction data.  The degenerate behaviour between the shape of the size-mass relation and any host mass dependence to the GC mass function is clear - and further observations of both quantities will be important in the dwarf galaxy regime.

As an example, in the right panel \ref{fig:f5} we show $R_{e}-M_{gal}$ relations from two literature studies focusing on higher mass galaxies.  The \cite{vanderWel14} studies yield nucleation fraction models which show a reasonable agreement with observed $f_{nuc}$ of $M_{gal} \geq 10^{9}$ galaxies.  However, both predict high nucleation fractions at low masses if extrapolated beyond their sample host mass limits.  Extrapolation of the curved \cite{Graham06} would favour a weakly varying GC mass function in order to reproduce the nucleation fraction - however the constraints from the heterogeneous data compilation shown are not sufficient to verify if such an extrapolation is accurate.

Figure \ref{fig:f5} suggests the host structural and GCMF scaling relations for a population of galaxies are crucial in modulating the efficiency of in-falling GCs, and setting the shape of the $f_{nuc}-M_{gal}$ distribution.  At face value it appears that that both \textit{the location and sharpness of the inflection point in the galaxy size-mass relation, and strength of any mass dependence of GCMF parameters} are important factors in setting the nucleation fraction in our model. In Appendix \ref{app:c} we further illustrate the impact with generalized size-mass relations.

\subsubsection{Dependence on GC formation distance}
Figure \ref{fig:f3i} shows our baseline model with fixed GC age $t_{form}=13$ Gyrs, and a variation in the GC formation location in each galaxy relative to the galaxy effective radius, $\zeta = R_{i}/R_{e}$.
Unsurprisingly this has a larger effect - with smaller formation distances producing higher nucleation fractions at a wider range of galaxy masses.  This results from the GCs having a proportionally smaller distance to successfully travel and survive.  The GC formation distance may change the width of the nucleation fraction distribution, but does not appear to alter the location of the peak value. We note that the distribution of GC formation distances  may be unrelated to the present day stellar density profile - especially as GCs may form from regions biased towards high gas pressure, and the total baryonic structure of the galaxies at high redshift may be much more chaotic.  Our stochastic model below explicitly allows for and presents variations in this parameter to address these effects.  

\subsection{Stochastic model}
After exploration of some parameters in our model we next wish to understand how typical variations in these properties, and other galaxy properties, manifest themselves in observational parameter spaces of interest ($f_{nuc} - M_{gal}, M_{NSC} - M_{gal}, M_{GC,obs} - M_{gal}$).  To do this we simultaneously incorporate variations in the GC mass function, initial GC formation distance, dissolution mass and galaxy size-mass relation into a stochastic, canonical model. To this we add scatter in the host galaxy Sersic index and DM fractions. We stress this is not meant to be a unique or optimal model, but rather illustrates the expected scale of variations due to stochasticity in some of these factors in real galaxy populations - especially in the low mass regime where shot noise in the GC populations may produce large variations.

We uniformly sample $10^5$ galaxies in log stellar mass over the range $4 \leq {\rm log M_{*,gal}} \leq 12$, and assign them a virial mass following the stellar-to-halo mass relation of \cite{Leauthaud12} with a scatter of 0.3 dex.  We impart a scatter of $\sigma_{n} \sim 1$ in the Sersic index - host mass, and $\sigma_{\rm log R_{e}} \sim 0.25$ dex in effective radius at fixed mass  The underlying size-mass relation at the low mass end spans the range shown in Figure \ref{fig:f5}.  For reproducing the nucleation fraction data we allow for correlated variations in the mass dependence of GCMF parameters ($\sigma, \mu$) comparable to the scales shown in the middle panels of Figure \ref{fig:f5}.  The fraction of star formation occurring in star clusters $\eta$ is uniformly sampled in log space between $-3 \leq \rm{log} ~\eta = -0.5$.  For predicting the total NSC and GC system mass we adopt a constant $\alpha = -2$ power law mass distribution for the GC populations, but allow variation in the upper cutoff mass to be the smaller of $M_{GC,max} = ${\rm min}[$10^{7}$,~$\eta M_{gal}$].  The GCs are randomly formed in each galaxy at a distance uniformly sampled between $0.2 \leq R_{i}/R_{e} \leq 0.8$.

\begin{figure}
\includegraphics[width=0.48\textwidth]{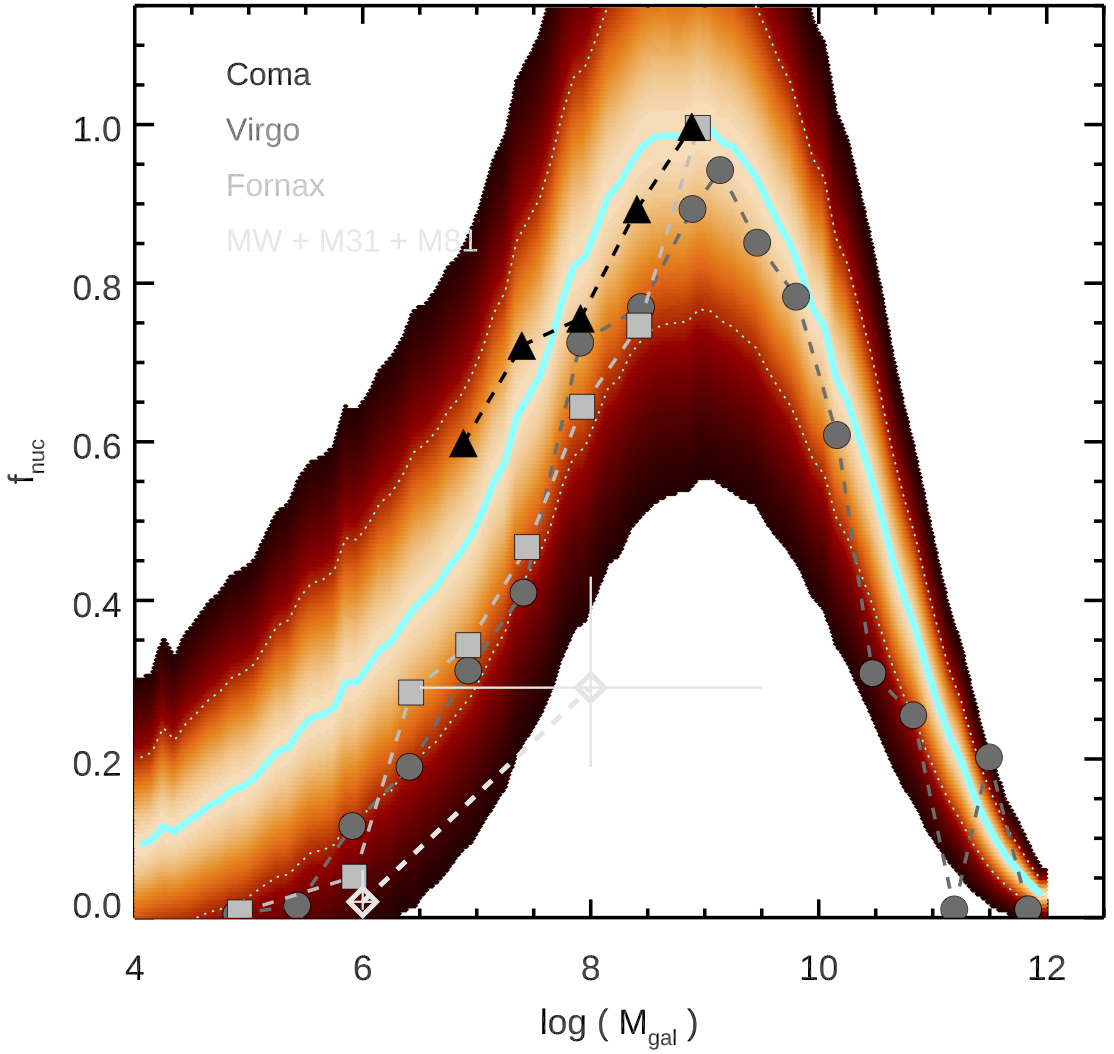}
\caption{Density distribution for $f_{nuc}$ predicted in a stochastic realization of our model.  Solid and dotted cyan line mean and $1-\sigma$ range for the mock galaxy populations predicted in our model when allowing variations in the model ingredients, as described in the text. Observational data are shown as in Figure \ref{fig:f2}.}
\label{fig:nucopt}
\end{figure}

Figure \ref{fig:nucopt} shows the median and $1-\sigma$ range of the nucleation fractions computed for this stochastic mock galaxy population.  While this is is not a fit to the data, the peak position and shape is well reproduced owing primarily to a combination of the size-mass relation shape and amount of mass dependence of the GCMF that we sample over.  The environmental dependence of the nucleation fraction is still not encompassed in this exercise, and is discussed further in Section 5.3.  We stress that this exercise is not meant to argue for a single process or solution - the factors we are discussing produce non-unique solutions.  For example a change in the size-mass relation could be counter-acted by changes to the GC mass spectrum or starting position.  The current plot simply takes into account many of the expected variations in galaxy properties with host galaxy mass, and illustrates how these may propagate through to variations in the nucleation fraction.  In Section 5 we will explicitly discuss potential impacts outside of these model ingredients.  A key advantage of this model will be whether the same model which reproduces the nucleation fraction, can also produce NSC and GC system mass estimates in agreement with observations.

\subsection{NSC and GC system masses}
Using Equations 10-13 we can predict the upper and lower limits, as well as expected average mass in a galaxy's NSC and surviving GC system.  In this section we explore how the predicted distributions compare to the observational data, and in the appendix how those model envelopes depend on a few key parameters.

\begin{figure}
\includegraphics[width=0.48\textwidth]{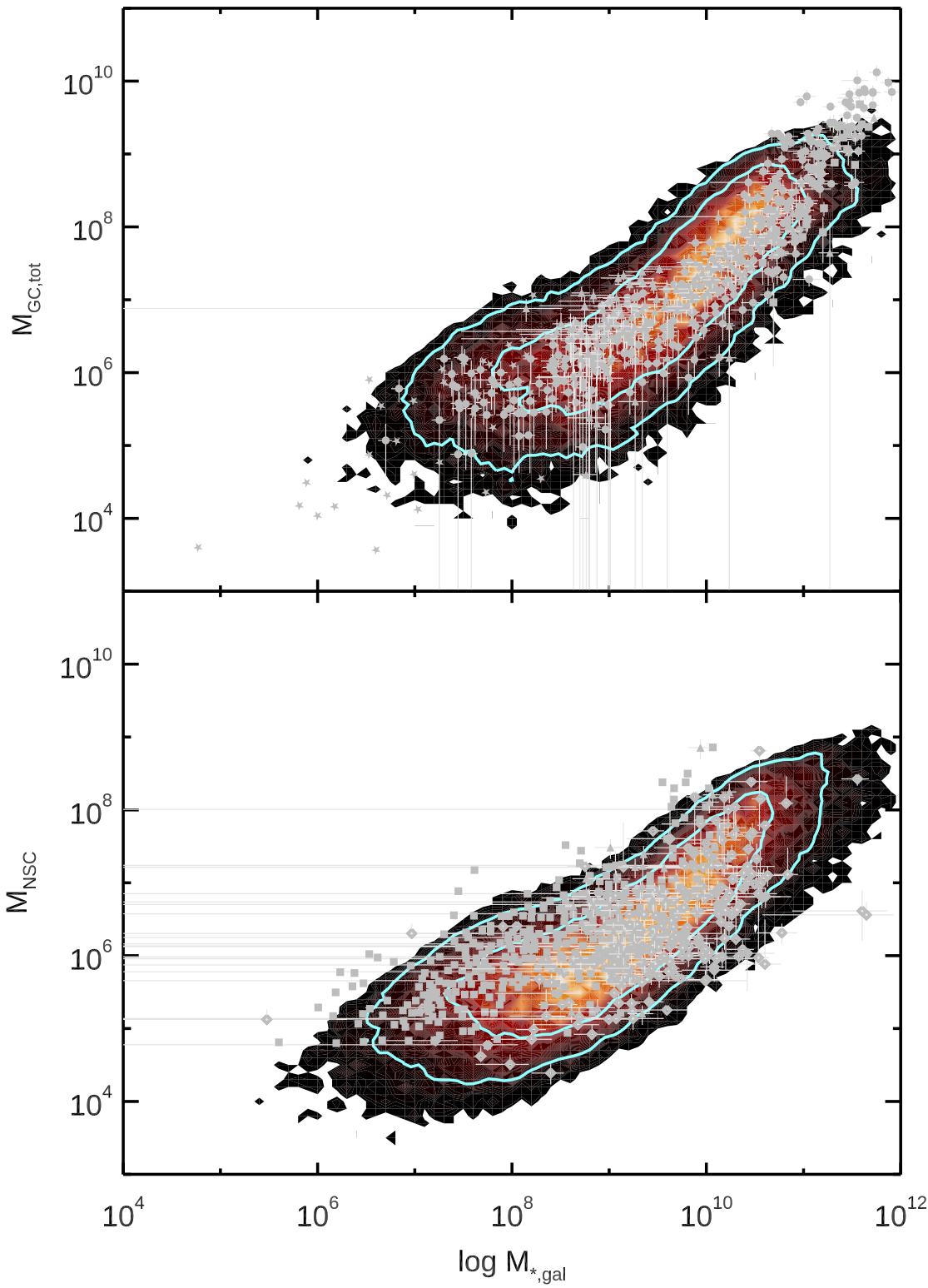}
\caption{Predicted mass in surviving GCs (\textit{top}) and mass of the host galaxy's NSC (\textit{bottom}) as a function of galaxy stellar mass.  Observational data are shown in grey (\textit{see text for references}).  Our canonical stochastic model predictions are shown as the red density distributions, with cyan lines indicating the parameter space containing $1,2-\sigma$ of the predicted distributions.}
\label{fig:obsopt}
\end{figure}

Figure \ref{fig:obsopt} shows the predictions for the total mass in GCs and the NSC mass for the same stochastic model described in the preceding section.  The density distribution of the mock galaxy populations cover the large intrinsic scatter of both relations and recover the shape of those scaling relations.

For the total GC system mass, there is an overabundance of observed high mass ($M_{gal} \gtrsim 10^{11}$) galaxies with large ($M_{GC,tot} \gtrsim 10^{9}$) GC system masses compared to our model.  This is expected as we do not consider galaxy mergers in this simple model.  At these galaxy masses, a significant fraction of the stellar mass and GC populations may be ex-situ and accreted during the course of the galaxy's hierarchical assembly (e.g., \citealt{Beasley18}).   As discussed by \cite{Elbadry19} the shape of the high mass $M_{GC,tot} - M_{gal}$ relation can be the result of this merging processes, with galaxies evolving along the sequence after many merger events.  It is therefore reasonable that we are not reproducing the GC mass populations for these most massive galaxies with the highest accretion fractions \citep{Boecker19}, given that we do not present galaxy merging here.  We have verified that including representative cosmological merger histories produce an elongation of our contours along the observed sequence and recovers these populations.

For the NSC mass scaling relation, the stochastic model distribution covers the large intrinsic scatter and mass range seen in the observations.  There is a minority population of dwarf galaxies with $M_{gal} \lesssim 10^{8}$  which have NSC masses larger than the contours predicted by our stochastic model (though consistent within uncertainties).  However as shown in Figure \ref{fig:obseta} extending our stochastic model to include clustered star formation efficiencies $log ~\eta \geq -0.5$ would produce models which cover these. The population of MW mass galaxies with extremely massive NSCs ($M_{NSC} \gtrsim 10^{8.5}$) may also be encompassed with an increased cluster formation efficiency - however these systems likely represent NSCs where additional mass growth occurs through \textit{in-situ} star formation.  In Section 5.4 we discuss how constrained reproduction of the NSC:GC system mass ratio may provide a way to constrain the \textit{in-situ} mass fraction of NSCs in individual galaxies. We note that it is possible that both of these populations of galaxies with over-massive NSCs could also have undergone host mass loss due to stripping in the cluster environment - shifting them to the left on the $M_{NSC} - M_{gal}$ plot.  This will be discussed further in Section 5.3.

The impact of additional parameters on our model's predictions for the NSC and GC system mass scaling relations are shown and discussed in detail in Appendix \ref{app:c}.  Of note is the dependence on galaxy size at fixed mass shown in Figure C8 in setting the NSC mass.  In particular at MW mass and above, the expectation value for the NSC mass decreases  for more extended galaxies.  This would provide a natural explanation for the observed difference in $M_{gal}-M_{NSC}$ relations reported in \cite{Georgiev16}.  If late type galaxies were slightly more extended than the ETG sample, a shallower slope as observed may be produced.  Relevant to the lowest mass galaxies, we note that reducing the initial formation distance $R_{i}/R_{e}$ or decreasing $R_{e}$ would both work to reduce the lower bound of the model envelope for $M_{NSC}$ and $M_{GC,tot}$, and hence can control the amount of intrinsic scatter at fixed galaxy mass (as the upper limit is always set by $\eta M_{gal}$).

\section{Discussion}

\subsection{On the correlation between \texorpdfstring{$M_{NSC} - M_{GC,obs}$}{Mnsc - Mgcobs}}
RSJ19 showed that the total mass in observed GCs is approximately equal to the NSC mass in galaxies which have both nuclei and GC systems in their sample.  At first glance this may appear to be a peculiar coincidence - why should the exact amount of GCs which form and survive have a close link to the NSC mass?  However we illustrate below why this is expected in a model where the NSC growth is primarily driven by star cluster in-spiral and mergers.

\begin{figure}
\includegraphics[width=0.48\textwidth]{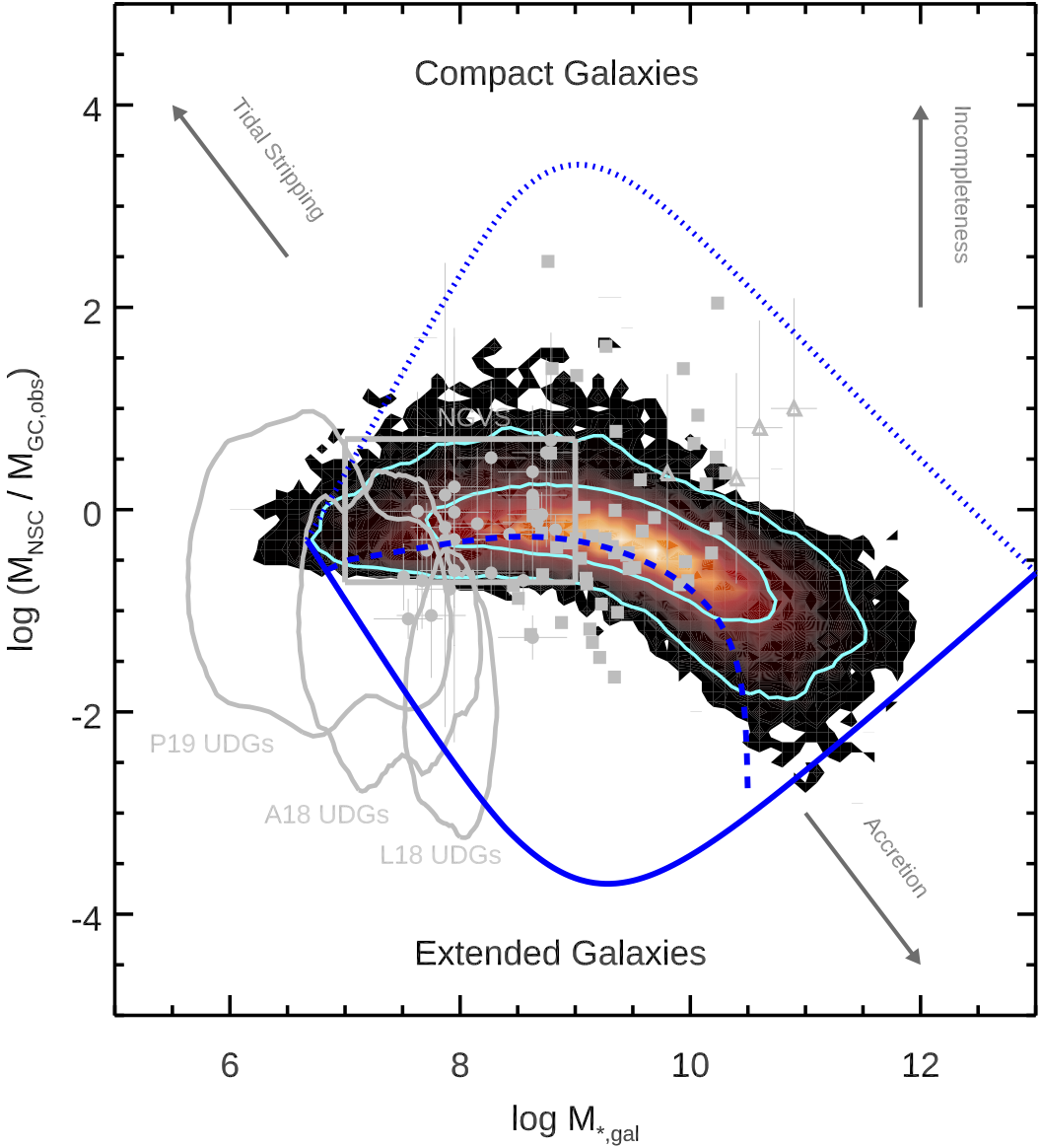}
\caption{Density distribution and cyan contours show stochastic model predictions for the ratio of NSC to GC system mass as a function of host galaxy stellar mass.  Analytic lower and upper limits for a single choice of model parameters are shown as the solid and dotted blue lines, and the average value as the dashed blue line.  Individual field and cluster galaxies (\textit{grey points}) and the range for galaxies in the NGVS survey (\textit{grey box}) exhibit $\sim 1$ dex scatter about a value of unity.  Host density dependence of our model predicts lower mass ratios for more extended galaxies - consistent with the position of populations of UDG/LSBs (\textit{grey contours}). Dark grey arrows indicate the impact of external astrophysical and observational effects on the data.}
\label{fig:rat}
\end{figure}

If we consider the limits on the NSC mass and GC system masses in \S 3.3, the bounds on the ratio of the two take the limits of:
\begin{equation}
\frac{M_{GC,lim}}{\eta M_{*,gal}} \leq \frac{M_{NSC}}{M_{GC,obs}} \leq \frac{\eta M_{*,gal}}{M_{GC,lim}\left(1 + {\rm ln}\left(\frac{M_{GC,lim}}{M_{diss}}\right)\right)}
\end{equation}
We show these limits on the ratio $M_{NSC}/M_{GC,obs}$ for a value of $\eta = 0.5$ as blue lines in the Figure \ref{fig:rat}.  The observational data from the Virgo and Coma clusters (RSJ, \citealt{Peng06,denBrok14}, as well as UDG studies \citep{Lim18,Amorisco18,Prole19} fall within the hard limits imposed by our analytic arguments. 

By considering the ratio of the expectation values for the two quantities we can understand whether the ratio of the NSC to total GC mass should be host mass independent and close to unity:
\begin{equation}
    \left\langle\frac{M_{NSC}}{M_{GC,obs}}\right\rangle = \frac{\left[\frac{1 + {\rm ln}\frac{M_{cl,max}}{M_{GC,lim}}}{1 + {\rm ln}\frac{M_{cl,max}}{M_{cl,min}}}\right]}{\left(1 - \frac{1 + {\rm ln}\frac{M_{cl,max}}{M_{GC,lin}}}{1 + {\rm ln}\frac{M_{cl,max}}{M_{cl,min}}} - \frac{M_{diss}}{\eta M_{*,gal}}\left(1 + {\rm ln}\frac{M_{diss}}{M_{cl,min}}\right)\right)}
\end{equation}


The contours from the same stochastic model in Figure \ref{fig:nucopt} and \ref{fig:obsopt} which trace this expectation value show good agreement with the bulk of the galaxy data.  However, why might the ratio of $M_{NSC}/M_{GC,obs}$ be close to unity in galaxies which have NSCs?

In the regime where NSCs and GC systems are well measured and both present in galaxies, a more intuitive understanding of the expected ratio of the mass in NSCs to surviving GCs can be gleaned from considering the average of the maximum and minimum values shown in Equation 14:
\begin{equation}
    \left(\frac{M_{NSC}}{M_{GC,obs}}\right)_{avg} = \frac{M_{GC,lim}}{2\eta M_{*,gal}} + \frac{\eta M_{*,gal}}{2M_{GC,lim}}\left(1 + {\rm ln}\frac{M_{GC,lim}}{M_{diss}}\right)^{-1}
\end{equation}

This average value of the NSC to GC system mass ratio is plotted in Figure \ref{fig:rat} as the blue dashed line.  It shows only a mild mass dependence in the range where galaxies are observed to have both NSCs and GC systems ($10^{7} \lesssim M_{*,gal} \lesssim 10^{10.5}$). At higher masses the $(\eta M_{*,gal})^{-1}$ term in the denominator of Equation 16 begins to dominate and drives the expectation value to more GC dominated regimes.

From this we see that on average:
\begin{itemize}
    \item The cluster formation efficiency can only increase the dispersion of the ratio, not the absolute value
    \item The galaxy mass dependence approximately cancels out
    \item Increasing the threshold mass for GC in-spiral decreases the ratio
    \item Increasing the threshold mass for evaporative dissolution increases the ratio
\end{itemize}
The host mass and cluster formation efficiency \textit{independence} of the ratio in this intermediate mass galaxy regime, is primarily a consequence of the most massive GC in the total population also being the most massive one which contributes to the NSC.  In other words,  the GC sub-populations which survive, and those which build up the NSC, are drawn from the same parent power-law distribution - and hence changing the upper limit $\eta M_{*,gal}$ does not alter the ratio of the two quantities.

A host galaxy mass dependence will enter if $M_{GC,lim}$ or $M_{GC,diss}$ is a strong function of $M_{*,gal}$, however the effect will be of order  $\mathcal{O} \sim \partial{\rm ln}M_{GC}/\partial{\rm ln}M_{*,gal}$. The GC dissolution mass is not expected to vary strongly with host mass, as a simple tidal argument would suggest:
\begin{equation}
\frac{\partial{\rm ln}M_{diss}}{\partial{\rm ln}M_{*,gal}} \propto \frac{3}{2}\left(\frac{r_{tid,GC}}{R_{e,gal}}\right)^{3}    
\end{equation}
which goes to a constant at the moment of disruption (e.g., when $r_{tid,GC} = D_{GC,gal} = R_{e,gal}$).  

Our limiting mass for in-spiral, $M_{GC,lim}$, is predominantly set by the galaxy density profile, and scales roughly as $M_{GC,lim} \propto R_{e,gal}^{2}$.  Any host mass dependence to the ratio of $M_{NSC}/M_{GC,obs}$ will then enter due to this change as: 
\begin{equation}
    \frac{\partial{\rm ln}M_{GC,lim}}{\partial{\rm ln}M_{*,gal}} = 2\frac{\partial{\rm ln}R_{e,gal}}{\partial{\rm ln}M_{*,gal}}
\end{equation}
For our canonical model using a size-mass relation based on \cite{Graham06}, this results in a change of at most a factor of 6 in the ratio.

Returning to the results in Figure \ref{fig:rat}, we see that galaxies at fixed stellar mass which have higher relative NSC masses, may plausibly have lower limiting masses for GC in-spiral ($M_{GC,lim}$).  Given that $M_{GC,lim} \propto R_{e}^{2}$, \textit{a prediction would be that on average, galaxies with values of $M_{NSC} \geq M_{GC,obs}$ have smaller effective radii at fixed stellar mass}, than galaxies with $M_{NSC} \leq M_{GC,obs}$.

Indeed we see that the UDG/LSB populations in Figure \ref{fig:rat} tend to be GC dominated systems - consistent with this line of argument.  The trend of decreasing values of $M_{NSC}/M_{GC,obs}$ across the \cite{Prole19}, \cite{Amorisco18}, and \cite{Lim18} samples is consistent with the increasingly strict size and density thresholds for the UDG sample definitions in those studies. We note \cite{Lim18} and \cite{denBrok14}  found that the Coma cluster dwarfs which were nucleated tended to have smaller sizes, or be more compact (larger Sersic indices). 

For now our model is suggestive that the connection between the observed GC system mass and NSC mass is the result of a common star cluster threshold mass for in-spiral and survival.  This in and of itself is a strong indication that NSC formation has important contributions from GC in-spiral.  Future work looking for correlations in galaxies' structural properties and their relative mass in GCs and NSCs, in a range of environments, will be valuable in further understanding the fate of bound star clusters in galaxies - and refine what amount of \textit{in-situ} NSC growth could be allowed within these observational constraints.  As an example Appendix \ref{app:c} presents explicit examples of how the expectation values change due to formation distance of the GC or changes to the host galaxy size-mass relation or cluster formation efficiency.

Other processes out of the scope of our model can change a galaxy's position in the parameter space of Figure \ref{fig:rat} - notably tidal stripping, which would work to reduce the number of GCs while leaving the NSC mass un-altered.  In Figure \ref{fig:rat} we have indicated how this would shift galaxies upwards (potentially also to the left if the tidal stripping reached the stellar body of the host) on this diagram.  Galaxies which undergo rich accretion histories would be expected to gain stellar mass, as well as accrete GCs.  These objects would not be accounted for in our model and should result in GC dominated systems preferentially in the high galaxy mass regime ($M_{*,gal} \geq 10^{11}$) where ex-situ fractions can exceed 50\% of the total mass of the galaxy.  Similarly, galaxies imaged around a small central field of view may have many of their GCs un-accounted for, and correction for this effect would shift galaxies upwards in Figure \ref{fig:rat}. However at larger radii a higher fraction of GCs are expected to be accreted, so this will not be a simple correction to account for.  We suggest that all these comparisons would be best done after separating (chemically, dynamically) the GCs most likely to be born in-situ in the host galaxy. For the interested reader, Appendix D shows explicit accounting of ex-situ contributions to the GC populations in our model.

\subsection{Interpretation of the shape of the \texorpdfstring{$f_{nuc}-M_{gal}$}{fnuc-mgal} relation}
While there are modulations in the width of the $f_{nuc} - M_{gal}$ relation, due to the initial distance the GC formed at in the galaxy or the shape of the GC initial mass function, Figures \ref{fig:f5} and \ref{fig:f6var} unequivocally demonstrate that a galaxy's size and structure play an important role in how efficiently GCs can contribute to formation of an NSC.

The observed nucleation fraction trend presented by RSJ19 requires an inflection in the galaxy size-mass relation at $M_{gal} \sim 10^{9-10}$ for our model to reproduce it.  The interpretation of the decreasing nucleation fraction at galactic masses above and below this, is that the in-spiral and tidal survival criteria for GCs become increasingly strict.  In both regimes the GCs start at \emph{proportionally} larger distances, and thus have a longer timescale to experience tidally driven evaporation and infall.

While our simple model is not a complete description of all processes expected to contribute to NSC formation, it suggests that the drop in nucleation fraction at low and high galaxy masses does not need to invoke inefficient {GC formation} or interactions with SMBHs, respectively.

This is given indirect support when comparing the observed mass budget in total GCs in galaxies, versus the observed NSC mass in those galaxies.   The data in Figure \ref{fig:obsopt} show that at all masses where NSCs are observed, there is enough mass in the galaxy's \emph{present day surviving} GCs to have made up the NSC mass budget.  At galaxy masses of $10^{8}$ for example, the typical NSC mass is about at the peak of the GC mass function ($5\times10^5$), suggesting only one massive GC may be needed to form the NSC.  

For comparison, from Equation 7,8 and 10 we see that if a galaxy has formed a massive enough star cluster to form an NSC, its integrated \textit{total} GCs ever formed will necessarily be larger than the NSC expectation value:
\begin{equation}
    \frac{M_{NSC}}{M_{GC,tot}} \leq \frac{1 + {\rm ln}\left(\frac{M_{cl,max}}{M_{GC,lim}}\right)}{1 + {\rm ln}\left(\frac{M_{cl,max}}{M_{cl,min}}\right)}
\end{equation}
For a power law distribution the integral over the full limits of the cluster mass range will always be larger than the sub-range defined by $\cup [M_{GC,lim},M_{cl,max}]$ in the numerator of the above equation.

This would suggest the \textit{total} mass in GCs ever formed will always exceed the final NSC mass, and that the drop in nucleation fraction at low galaxy masses is not due to insufficient formation of GCs (as the galaxies also show nuclei).  This is different than the mass ratio of the observed mass in \textit{surviving} GCs ($M_{GC,obs}$), which can be smaller or larger than the NSC mass and was discussed in \S 5.1.
The low nucleation fractions below $M_{*,gal} \lesssim 10^{9}$ are then likely not due to a GC formation inefficiency or mass-budget issue, but rather that the conditions for survival and infall become increasingly less favourable due to the proportionally larger sizes of low-mass galaxies.

We stress that our model is an incomplete description of the processes expected to occur over the lifetime of the NSC.  However it is flexible enough to allow for these to still occur.  For example, the recovered nucleation fraction curve could be over-produced at the high mass end if the size-mass relation were changed in our model.  

Figure \ref{fig:f5} shows this explicitly as the black dashed line, which represents an evolution of the sizes back to $z \sim 2$ following \cite{vanderWel14}.  If the GC in-spiral predominantly occurred at this epoch when the galaxies were more compact, we see that our model predicts a broader nucleation fraction distribution.  At the low mass galaxy end, this could provide another interpretation of the environmental offset in nucleation fraction - with nuclei of dwarf galaxies in high density environments forming earlier due to the high density environments collapsing first and their galaxies quenching early (See $\S 5.3$).  

At the high mass end this broader distribution allows for processes such as binary SMBH mergers to destroy NSCs and further reduce the nucleation fraction \citep{Antonini}.  Taken at face value such a process would need to reduce the number of nucleated galaxies in this example by less than a factor of two. The flexibility of our model and its link to the field star cluster populations will hopefully provide a way to better assess the relative contributions of such additional processes in the future.

\subsection{Impact of environment on the host galaxy star cluster relations}

The data compilation of RSJ19 shows a clear second-order tendency for the nucleation fraction at fixed galaxy mass to be larger in denser environments (from MW/M31/M81 through Virgo/Fornax to Coma). The distribution predicted from our stochastic model does not fully encompass this variation at fixed host mass (Fig. \ref{fig:nucopt}), suggesting other factors may be responsible for this environmental dependence.  While tempting to associate the higher $f_{nuc}$ in denser environments to tidal stripping of the host galaxy, the galaxy populations in those clusters are all observed to have GCs around them, which would argue against significant tidal evolution for the host galaxies \citep{Smith15}.

Instead the higher $f_{nuc}$ values for Coma cluster galaxies below $M_{*,gal} \lesssim 10^{9}$ may be a consequence of observational selection effects and/or complete {disruption} of the low density galaxies in a cluster.  We have already seen that nucleation is more efficient for more compact galaxies (Figure \ref{fig:f5}).  These compact galaxy populations may be \textit{preferentially surviving and/or detected} in the observational sample, relative to more diffuse galaxies (which are less likely to be nucleated in our model).  Rather than an intrinsic difference in nucleation efficiency, the observed environmental dependence to the $f_{nuc}$ data may then result from the low density local volume environments preserving (and/or detecting) more faint, diffuse galaxies (with their typical lack of nuclei).

\begin{figure*}
\includegraphics[width=0.78\textwidth]{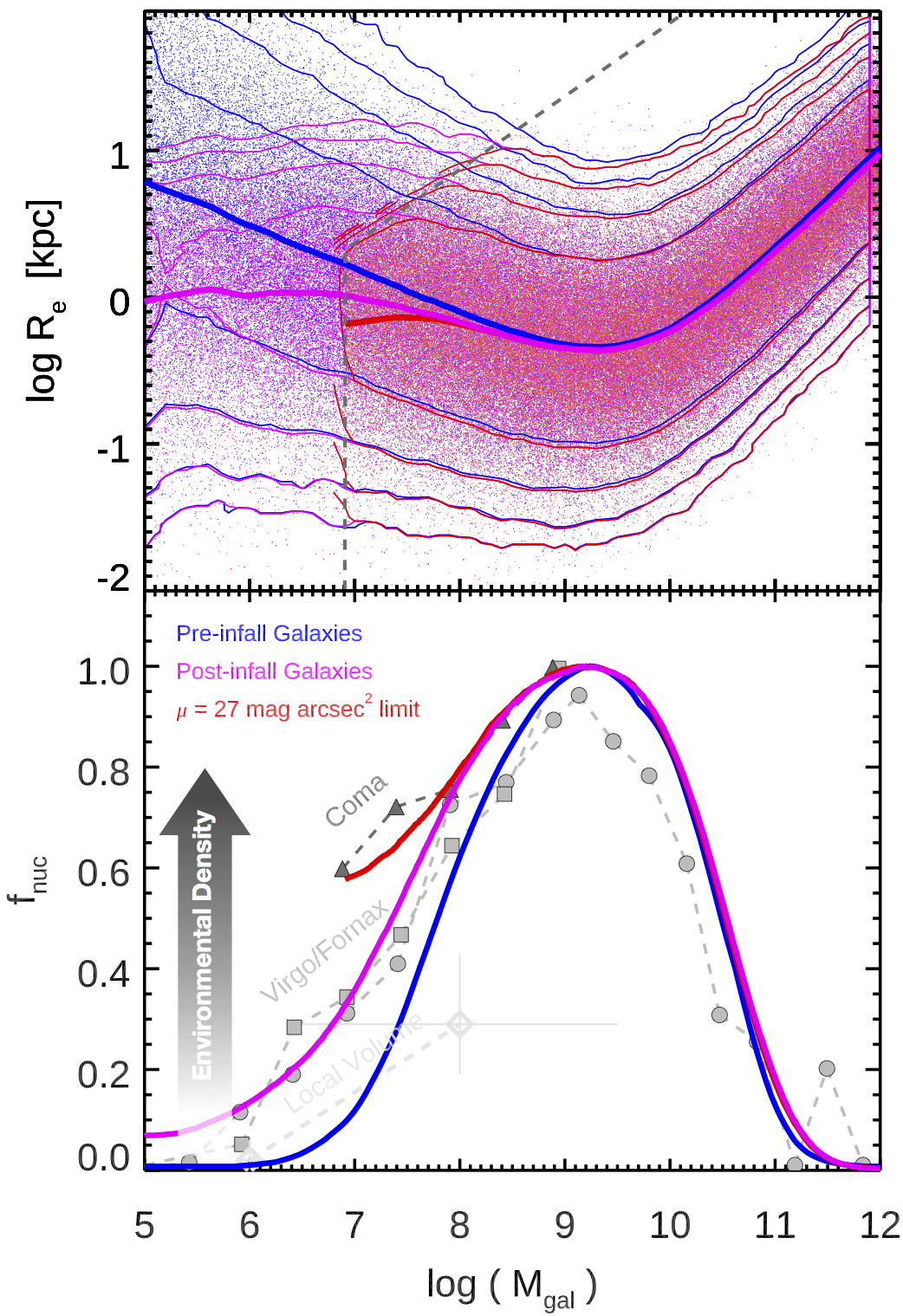}
\caption{Initial population of galaxies in a Coma-like cluster environment (\textit{blue}), after orbital evolution in the cluster (\textit{magenta}), and with a observational detection threshold (\textit{red}).  The preferential destruction of the most diffuse galaxies in the cluster, leads to a environment dependent change in the nucleation fraction (\textit{bottom}) which qualitatively tracks the dependence in observational data.}
\label{fig:coma}
\end{figure*}

To illustrate the effect of the environment altering a \textit{surviving} population of galaxies, we stochastically generate a population of galaxies which initially follow a canonical size-mass relation presented in $\S 4$.  These are shown as the blue dots and contours in Figure \ref{fig:coma}.  We then compute the likely distribution of galaxy orbital pericentres in a cluster environment by sampling the distribution of $D_{per}$ from the cosmological simulations of \cite{Wetzel11}.  These $D_{per}$ distributions are dependent on the galaxy cluster mass and virial radius, and we stochastically draw a mass for a Coma-like cluster between $M_{NFW,Coma} = 1-4\times10^{15}$, and adopt $R_{vir,Coma} = 2.9$ Mpc.  

Each galaxy of stellar mass $M_{*,gal}$ is represented as an NFW halo with virial mass prescribed from the stellar-to-halo mass relation of \cite{Leauthaud12}, and DM halo concentration from the redshift zero halo mass-concentration relation of \cite{Dutton14}.  For each galaxy, we compute its enclosed mass within the half-light radius ($M_{enc,gal} = M_{NFW,gal}(<R_{e}) + M_{*,gal}/2$), and the enclosed cluster mass that the galaxy would see at it's orbital pericentre in the cluster ($M_{enc,Coma} = M_{NFW,Coma}(<D_{per,gal})$.  A conservative survival limit for the cluster population then is computed by selecting galaxies which have effective radii less than:
\begin{equation}
    R_{e} \leq D_{per,gal}\left(\frac{M_{enc,gal}}{M_{enc,Coma}}\right)^{1/3}
\end{equation}

The preferential destruction of the most diffuse galaxies is visible in the top panel of Figure \ref{fig:coma}. The corresponding impact on the nucleation fraction is to leave a population galaxies which are on average more compact, and thus tend to be on average more nucleated.  An additional threshold for surface brightness detection of the galaxy nuclei (here above the sky, or host galaxy light of $\mu = 27$ mag~ arcsec$^{-2}$) further increases the nucleation fraction at fixed galaxy mass for the same reason.

The changes to the nucleation fraction due to these effects span the variation at fixed galaxy mass seen in the observational data of RSJ19.  This effect suggests that regardless of when the galaxies are nucleated, \textit{the nucleation fraction will increase as the environmental density does}.  The prominent GC populations for these same galaxies again argues against tidal stripping as a mechanism to produce the environmental trend, especially as the gas reservoir for GC driven NSC formation would also be depleted.  Similarly, while Figure \ref{fig:f3i} shows higher nucleation fractions for GCs which formed initially closer to the galaxy centre, there is no reason this should correlate with present day galaxy environment at fixed host mass.  Rather this second order environmental trend of the nucleation fraction appears likely due to preferential survival and detection of the densest galaxies - which have internal structures most conducive to the formation and survival of GCs that in-spiral and build the NSCs. 

Could other factors partially contributed to this environmental dependence?  Figure \ref{fig:f4} demonstrated that changing the GC mass function also alters the width of the $f_{nuc} - M_{gal}$ relation.  There is observational evidence that the mean of the GC mass function may vary with {host galaxy mass}, but little direct evidence of such changes with present day extra-galactic environment.  An intrinsic mass dependence to the peak of the GC mass function, $< M_{GCMF} >~ \propto M_{gal}$, could explain the environmental dependence in $f_{nuc}$ if cluster galaxies were quenched sufficiently early. In this case they may have formed a GC population representative of larger galaxies (which they would have evolved to become if not for environmental quenching). In order to explain the nucleation fraction in Figure \ref{fig:f4}, the results on GC mass function variation from \cite{Jordan07} would naively require that the Coma cluster dwarfs have a deficit in $z=0$ galaxy stellar mass of $\sim 3.5$ dex relative to a field dwarf which formed stars over a Hubble time.  Given typical SFHs and stellar populations of these cluster dwarfs, this amount of mass difference is quite unlikely.

The nucleation fraction model presented Figure \ref{fig:f5} showed a broader distribution when using a $z=2$ size-mass relation, as the smaller galaxies at that time would be more conducive to nucleation if the GC in-spiral process occurred at that epoch.  Given that the cores of galaxy clusters form from some of the earliest/largest overdensities in the Universe, one could imagine their galaxies assembling earlier while the galaxies were more compact than dwarfs in the field at the same redshift \citep{vanderWel14}. The higher gas fractions and more turbulent and highly pressurized ISMs at these redshifts may also work to aid formation of massive GC populations - shifting the peak of the GC mass function to higher values and increasing the nucleation probability for the galaxies.  Subsequent quenching of these galaxies could additionally shift them to lower effective stellar masses at present day, increasing the nucleation fraction at fixed stellar mass.

While plausible, explaining the environmental trends in nucleation fraction with such offsets in NSC formation epochs is complicated by the lengthy assembly histories for the Virgo and Fornax clusters presented in RSJ19, which are still assembling via infall of sub-groups of galaxies.  Additionally the SFHs of dwarfs of various morphological types show that many were able to form significant and similar populations of massive clusters - suggesting that the required local ISM conditions at those epochs could be found in galaxies over many different global environments. 

Our favoured explanation for the nucleation fraction offset with environment is thus preferential destruction and non-detection of diffuse, non-nucleated dwarf galaxies in cluster environments.  Assessing additional contributions from variations in GC mass function or formation times could perhaps be done by examining the stellar populations of the NSCs.  For example, finding older and more extended NSCs in the cluster dwarfs compared to field dwarf galaxy NSCs would suggest some formation epoch difference could play a role.  Most useful will be to see how well all three observables ($f_{nuc}$, $M_{NSC}$, $M_{GC,obs}$) are reproduced for a tailored size-mass relation for a population of galaxies within a specific environment - as the NSC and GC system masses can also be influenced by differences in the GC mass function, or formation epochs as shown in Appendix \ref{app:c}.

An important application/test of our model in the future may be to ascertain if there are inconsistencies between the observed size-mass relation and nucleation fraction for a population of galaxies.  This could be the case if the GCs which form the NSCs were born and in-spiraled at early times (consistent with the predominantly old ages of NSCs); while the host galaxy structure seen today was subsequently altered by the environment of the galaxy cluster \textit{after} the NSC was assembled.  The nucleation fraction model would still be driven by the curved \emph{initial} size mass relation, and produce the nucleation fractions preserved in the galaxy population today.  While the host galaxies may have moved on the size-mass (and $M_{GC,obs}-M_{gal}$) planes due to tidal stripping or accretion driven size growth.  In this case \textit{discrepancies between the size-mass relation needed to fit the nucleation fraction data, and the observed present day sizes, may provide an estimate of the amount of time different galaxies have spent in the cluster environments}.

\subsection{The role of \textit{in-situ} star formation in NSC evolution}
Thus far we have considered that galaxy NSCs are built up entirely through in-spiral of GCs.  As shown in several studies (e.g., \citealt{Walcher06,Neumayer20}), there is direct evidence for populations of very young stars ($\sim 100$ Myrs) in many NSCs, suggesting that in-situ SF does occur (e.g., \citealt{FeldmeierKrause15,Alfaro19}).  While our model does not explicitly include mass growth due to in-situ star formation, the model framework predicts two complementary quantities which allow indirect estimation of the amount of an NSCs mass assembled through in-situ star formation.

For a given host galaxy mass and size, our model predicts the total mass in surviving GCs as well as the total mass of the NSC.  Thus, given an observed host galaxy size and mass, one can marginalize over a few free parameters ($\eta, M_{GC,max}, M_{diss}$) to assess how well the pure GC in-spiral model simultaneously reproduces the observed NSC and GC system mass.  For systems with strong amounts of \textit{in-situ} NSC growth a clear discrepancy in the predicted NSC mass for fixed GC system mass will be seen.  Appendix \ref{app:d} shows an example of how the in-situ fraction of the NSC depends on the GC system and galaxy properties.

In other words we suggest that the dominance of in-situ NSC growth can be examined in individual galaxies by assessing how well our model reproduces their ratio of star cluster to NSC masses (see Fahrion et al., in prep.). Further chemical predictions from our model will also aid in disentangling the relative contribution of \textit{in-situ} SF in nuclei given the wealth of upcoming high spatial resolution spectroscopic observations of NSCs (e.g., \citealt{Fahrion19})

A full exploration of how $f_{nuc}$ would vary in the case of a pure in-situ NSC model is beyond the scope of this paper. However, we note that recent simulation work by \cite{Guillard16} found that in-spiraling massive GCs can accrete gas from the host galaxy and either form stars during the in-fall, or bring this gas to the very central region of the proto-NSC.  Clearly further study is needed, but this would provide a self-consistent way to reconcile our NSC model with the presence of young NSC stellar populations.  

\subsection{Caveats and future outlook}
The presented model is straightforward, however we believe there is enough flexibility for it to provide testable predictions for key mechanisms driving NSC formation in galaxies.  In particular, given the observed variation seen in most GC and NSC scaling relations, the ability to describe the intrinsic spread of star cluster scaling relations can be considered an important success.  At present, accounting for only GC in-spiral, the model can reproduce much of the observational data available on nucleation fraction and NSC and GC system masses.  However we stress that the goal of this work is to see how far we can go with this process before we need additional ingredients.  None of the model curves we present are unique solutions, nor do we believe them to be complete descriptions of the processes responsible for the NSC scaling relations, nor the simplified host galaxy potentials.  However the straightforward framework of the model may still allow us to understand the baseline contribution from star clusters in-spiral, and how this varies with environment, host galaxy or star cluster properties.

Understanding the impact of additional mechanisms in excess of this baseline is the next step, and synergy with existing models in literature which treat these will provide a helpful way forward in addressing the wealth of observational data on NSCs.  For example, our chemical models for NSC formation (Leaman et al. in prep) already indicate some amount of \textit{in-situ} star formation will be required to reproduce the most metal rich stars in e.g., the MW NSC.  We anticipate that chemical predictions along with subsequent work on NSC size-mass relations, will let us further break the degeneracy between the fraction of \textit{in-situ} and GC based NSC growth, and aid in interpreting high spatial resolution spectroscopic observations of NSCs.

In addition to star formation from gas inflows to the NSC, we have not considered destructive processes from massive black holes, which are known occupy the centres of many galaxies/NSCs \citep{Georgiev16}.  For single SMBHs it is perhaps still unclear how large their impact on the NSCs is.  For example if the NSC and host potential are significantly triaxial, the stellar loss cone can be re-filled more efficiently by the centrophilic orbits permitted in those potentials \citep{Bockelmann02}.  However as shown in the models of \cite{Antonini} mergers leading to binary IMBHs/SMBHs are likely to have a far greater impact on the NSC.

In the highest mass galaxies where major mergers are more common our model could allow for this additional process. As seen in Figure \ref{fig:f5}, a mild change to the slope of the size-mass relation increases our nominal nucleation fraction, which would allow room for subsequent destruction of NSCs due to binary BH merger events.  A follow-up paper will show that our current model can still reproduce $M_{BH} - M_{gal}$ relations however, even if the NSCs in our model predate the BHs (c.f., \citealt{Bournaud07}).

The impact of accretion of star clusters from galaxy mergers is presented briefly in Appendix D - however studies in literature (e.g., \citealt{Elbadry19}) show that the consequence of this on the $M_{GC,obs} - M_{gal}$ relation is to evolve along the sequence.  As our model predicts the GC system and NSC properties of a host in absence of accretion, comparison to observations should either: self-consistently model the accretion history of the galaxy (e.g., \cite{Beasley18}, or compare these model predictions to a subset of GCs which are kinematically and chemically most likely to be formed in the host galaxy.

The advantage of this current model is that it serves as a flexible analytic baseline for the expectations of a single process.  The model can be supplemented by additional processes, but in its current form it still allows for an efficient characterization of the relative impact of a few key processes in NSC formation.  For example, as discussed in $\S 5.3$, predicting the nucleation fraction for a population of galaxies in a single environment using their observed galaxy size-mass relation can let us assess whether environmental processing or strong contributions from processes beyond GC in-spiral are required (e.g., SMBH driven destruction).  For individual galaxies, discrepancies between the predicted NSC and GC system masses will strongly constrain to what extent \textit{in-situ} SF has played a role in the NSCs formation. Application to observational data sets will be most useful when galaxies with star cluster masses as well as NSC sizes, chemistry and masses are available.  Such data sets are already starting to be assembled in large numbers for galaxies in nearby clusters (e.g., \citealt{Sarzi18,Lyubenova19})


\section{Conclusions}
We present a straightforward model for the nucleation probability of galaxies which assumes that the efficiency of NSC formation is linked to the in-spiral and survival efficiency of GCs in the host galaxy.  The host galaxy dependent limiting mass for GCs to contribute to the NSC build-up predicted in our model ($M_{GC,lim}$) shows excellent agreement with the observed most massive surviving GCs in local galaxies.   

We show that the strongest factor influencing whether a host galaxy will form an NSC, is the galaxy size at fixed stellar mass.  For the galaxy size-mass relation derived from \cite{Graham06}, which has an inflection point consistent with galaxy sizes in the local Universe, the model produces nucleation fractions in excellent agreement with results from the NGVS study of galaxies in four different cluster environments.

The efficiency of NSC formation in galaxies has a second order dependence on the form of the GC mass function, or birth location in the host galaxy, but the peak nucleation fraction for galaxies at $M_{gal} \sim 10^{9}$ stellar mass, appears as a direct result of those galaxies having the smallest effective radii.

A novelty of our model is that it simultaneously predicts the mass of the NSC and the surviving mass of GCs in the host galaxy.  We compare these to observations of galaxy GC system mass and NSC mass from literature and find that the shape and intrinsic scatter are reproduced by the same model which fits the nucleation fraction data.

The drop-off in $f_{nuc}$ at low galaxy masses, is therefore not due to an inability in forming high mass GCs, as observed galaxies in this mass range have formed sufficient mass in GCs, but rather that the mass threshold for successful GC in-spiral and survival is high because of the proportionally larger galaxy sizes.  The presented model has room to accommodate potential BH driven destruction of NSCs, but would only require reducing the number of nucleated high mass galaxies by less than a factor of two.

The roughly equal mass in NSCs and GC systems observed in galaxies is interpreted in our model as a result of a parent power-law mass distribution for clusters being split by a single mass threshold ($M_{GC,lim}$).  Star clusters above this contribute to the NSC mass, while those below contribute to the surviving GC population.  Variations in this ratio are bounded by the same upper limits for both populations however, resulting in an approximate host galaxy mass independence to the ratio of $M_{NSC}/M_{GC,obs}$.

Two predictions from our model, which appear consistent with available data are: 1) at fixed stellar mass, more compact galaxies should have more mass in their NSCs than GC systems, and 2) preferential tidal destruction and non-detection of the most diffuse galaxies in cluster environments results in a bias in the recovered nucleation fractions.  The second point follows directly from the first, as the most extended galaxies which are preferentially destroyed in the cluster environment are the least likely to be nucleated in our model - and the the combined effect results in higher nucleation fractions at fixed galaxy mass in denser environments.

While GC in-spiral is only one process contributing to galaxy nucleation, that our simple model can simultaneously reproduce the nucleation fraction and mass of the NSC and GC systems in galaxies, provides support that GC in-spiral is an important mechanism for NSC formation in galaxies in addition to in-situ SF.

\section*{Data Availability}
The underlying data which our analytic models are compared to are provided in the referenced works, and further requests should be directed to those authors. 

\section*{Acknowledgements}
We thank the anonymous referee for an extremely helpful report which greatly improved the manuscript. The authours thank Katja Fahrion, Nadine Neumayer, Allison Merritt, Mariya Lyubenova, Alessandra Mastrobuono-Battisti, Eric Emsellem, Morgan Fouesneau, Anna Sippel, Iskren Georgiev and Reynier Pelletier for insightful comments.  We additionally thank Rub\'{e}n S\'{a}nchez-Jannsen for providing the observational data for the model comparisons and additional encouragement and suggestions. RL thanks AH and the V\&I Cottage in Peebles for hospitality during formulation of this work.  This work was made possible with funding from the Natural Sciences and Engineering Research Council of Canada PDF award, DAAD
PPP project number 57316058 "Finding and exploiting accreted
star clusters in the Milky Way", and funding from the European Research Council (ERC) under the European Union's Horizon 2020 research and innovation programme under grant agreement No 724857 (Consolidator Grant ArcheoDyn)




\bibliographystyle{mnras}
\bibliography{fnuc} 




\appendix

\section{Integrated Mass loss During In-spiral}\label{app:a}
The fitting function for the evaporation rate in \cite{Madrid17} is given as:
\begin{eqnarray}
&&\xi(R,t) = a log_{10}(t) + b\\
&&a(R) = -10^{(3e^{-4log_{10}(R-1.05)}-1.2)}\nonumber\\
&&b(R) = 10^{(2.5e^{-4log_{10}(R-0.4)}-0.5)}\nonumber
\end{eqnarray}

With this, the integral on the left of Equation 6, which is proportional to the GC mass loss over the orbital infall, is:
\begin{eqnarray}
\Delta M_{GC} &=& \int_{R=0}^{R_{i}} \xi(R,t)R  dR = \\
 && 0.147 ~{\rm erfi}\left(\frac{2.628}{(R - 1.65)^{2}}\right){\rm log_{10}}(t_{form}) \nonumber\\
 && - 0.063\times10^{\frac{3}{(R - 1.65)^{4}}} \times \nonumber\\
 && \left(0.5 (R - 1.65)^{2} + 1.65 (R - 1.65)\right){\rm log_{10}}(t_{form}) \nonumber\\
 && -\frac{0.169~ {\rm log_{10}}(t) ~\Gamma\left\lbrace0.75, \frac{-6.908}{(R - 1.65)^4}\right\rbrace}{\left(\frac{-1}{(R - 1.65)^4}\right)^{3/4} (R - 1.65)^3}\nonumber\\
 && + 0.316\times10^{\frac{2.5}{(R - 0.4)^{4}}}(R - 0.4) \nonumber\\
 && + \frac{0.490 ~\Gamma\left\lbrace0.75, \frac{-5.757}{(R - 0.4)^{4}}\right\rbrace}{\left(\frac{-1}{(R - 0.4)^{4}}\right)^{3/4} (R - 0.4)^{3}}\nonumber
\end{eqnarray}
Here erfi is the imaginary error function, erfi$(x) = -i{\rm erf}(ix)$.  For positive $R$, our occurrence of the incomplete Gamma function, $\Gamma(\alpha,x)$, takes on a negative argument of x, which is not trivially evaluated.  It is likely to be a small factor however, as at least for values of $\alpha+1/2$ the x dependence drops out, leaving Re$\lbrace \Gamma(n + 1/2,-|z|)\rbrace = \Gamma(n+1/2)$ (Thompson, 2013).

To confirm the $R$ dependence of the argument in the incomplete Gamma function, we follow \"{O}z${\rm \underaccent{\cdot}{c}}$a\u{g} and Ege (2014) who extended the lower incomplete Gamma function to negative arguments, for arbitrary non-integer $\alpha$.
\begin{equation}
\gamma(\alpha,x_{-}) = \int_{0}^{-x} |u|^{\alpha -1}e^{-u}du
\end{equation}

For $\alpha = 3/4$ this integral becomes:
\begin{eqnarray}
&&\gamma(\alpha,x_{-}) = {\rm Re}\left\lbrace (x^{1/4} ({\rm sgn}(x) - 1) \Gamma(3/4, x))/(2 (-x)^{1/4}) \right. \\
&&\left. -\left(\frac{1}{4} + \frac{i}{4}\right) ({\rm sgn}(x) + 1) \left((1 - i) \Gamma(3/4, x) + \left(\sqrt(2) + (-1 + i)\right) \Gamma(3/4)\right)\right\rbrace\nonumber
\end{eqnarray}

For values of galactocentric distance, $1 \leq R/pc \leq 20000$ the two evaluations of $\Gamma$ in Equation A2 are bounded by $0.25379 \leq \Gamma \leq 0.75379$.

\section{Incorporating Observational Detection Limits in the Model}\label{app:b}
The analytic value $M_{GC,lim}$ for the GC mass-loss during in-spiral formally defines the limiting mass that will be lost for the GC to still arrive in the centre.  However this would leave the arriving GC with zero mass.  The practical value needs to be some factor $\gamma$ above this.  For our canonical model we have adopted a conservative limit of $M_{GC,lim} = 2M_{loss}$ as significant non-linear dissolution effects may occur if the GC has lost more than $50\%$ of its mass \citep{Webb14}.  
However we can also ask what the value of $\gamma$ should be in order to still detect a GC in the central region of a galaxy observationally.  This limit is set by either the GC magnitude relative to the sky brightness, or relative to the host galaxy light profile.  Both of these are likely to be lower than $\gamma = 2$.  As an example relevant to the Virgo cluster observations of SJ19, we describe the effect on $\gamma$ due to sky or galaxy brightness detection limits. The final limit corresponds to the case where the observational threshold for detecting a NSC in a galaxy of any  mass ($M_{NSC,lim}$) is:
\begin{equation}
    M_{GC,lim} = M_{loss} + M_{NSC,lim}
\end{equation}

For the case where the contribution from the host galaxy begins to dominate over the sky contribution we assume a functional form similar to RSJ19 of:
\begin{equation}
    M_{GC,lim} = M_{loss} + 10^{a{\rm exp}\left(\frac{\rm log_{10}}{b}\right) + c}
\end{equation}
For the NGVS survey limits in SJ19, a close match can be found for $(a,b,c) = 0.02, 2.8, 4.5$.

The change to the predicted galaxy nucleation fraction is indistinguishable for the case of the NGVS detection limits.  There is an equally small shift in the total GC system and NSC masses, however this is degenerate with the bound star cluster fraction, $\eta$.

\section{Additional parameters modifying the model predictions}\label{app:c}

\subsection{Effects on the nucleation fraction}
\subsubsection{GC formation distance}
\begin{figure}
\includegraphics[width=0.48\textwidth]{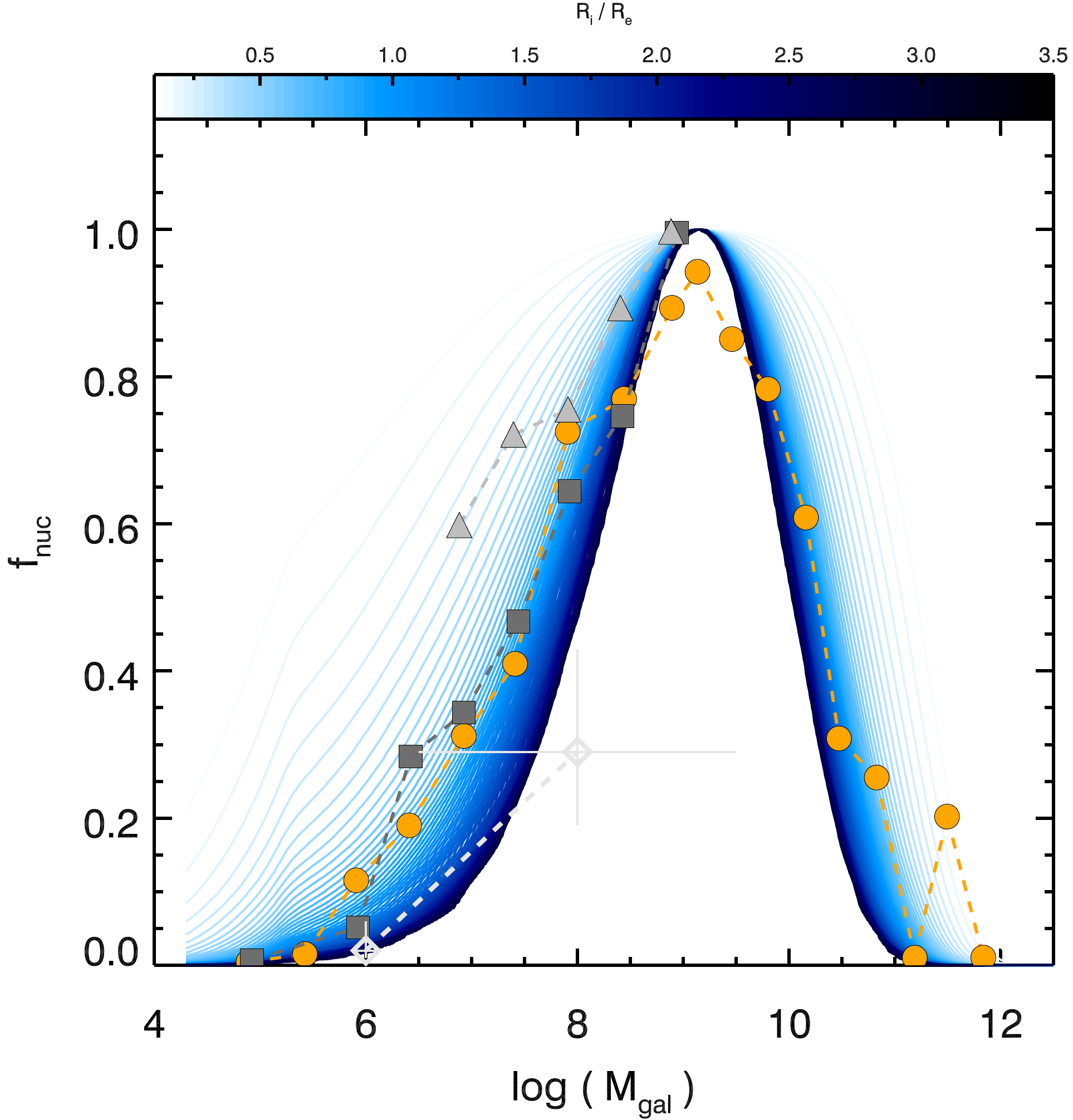}
\caption{Same as Figure 2, but models show the variation in starting distance for the GC infall, in units of the galaxy effective radii. GC age is held fixed at 13 Gyrs.}
\label{fig:f3i}
\end{figure}

In Figure \ref{fig:f3i} we show the impact in the nucleation fraction curve if the GCs formed at different factors of the host galaxy effective radius, $R_{i}/R_{e}$.  For smaller formation distances, we see that galaxies over a wider range of masses are more likely to be nucleated.  While we cannot formally seperate the two processes, the variations seen in this Figure is suggestive that the joint constraint $M_{GC,lim}$ in these regimes is not indicative of GC destruction efficiency, but rather inefficient GC in-spiral.

\subsubsection{Generalized size-mass relations}
\begin{figure*}
\includegraphics[width=0.98\textwidth]{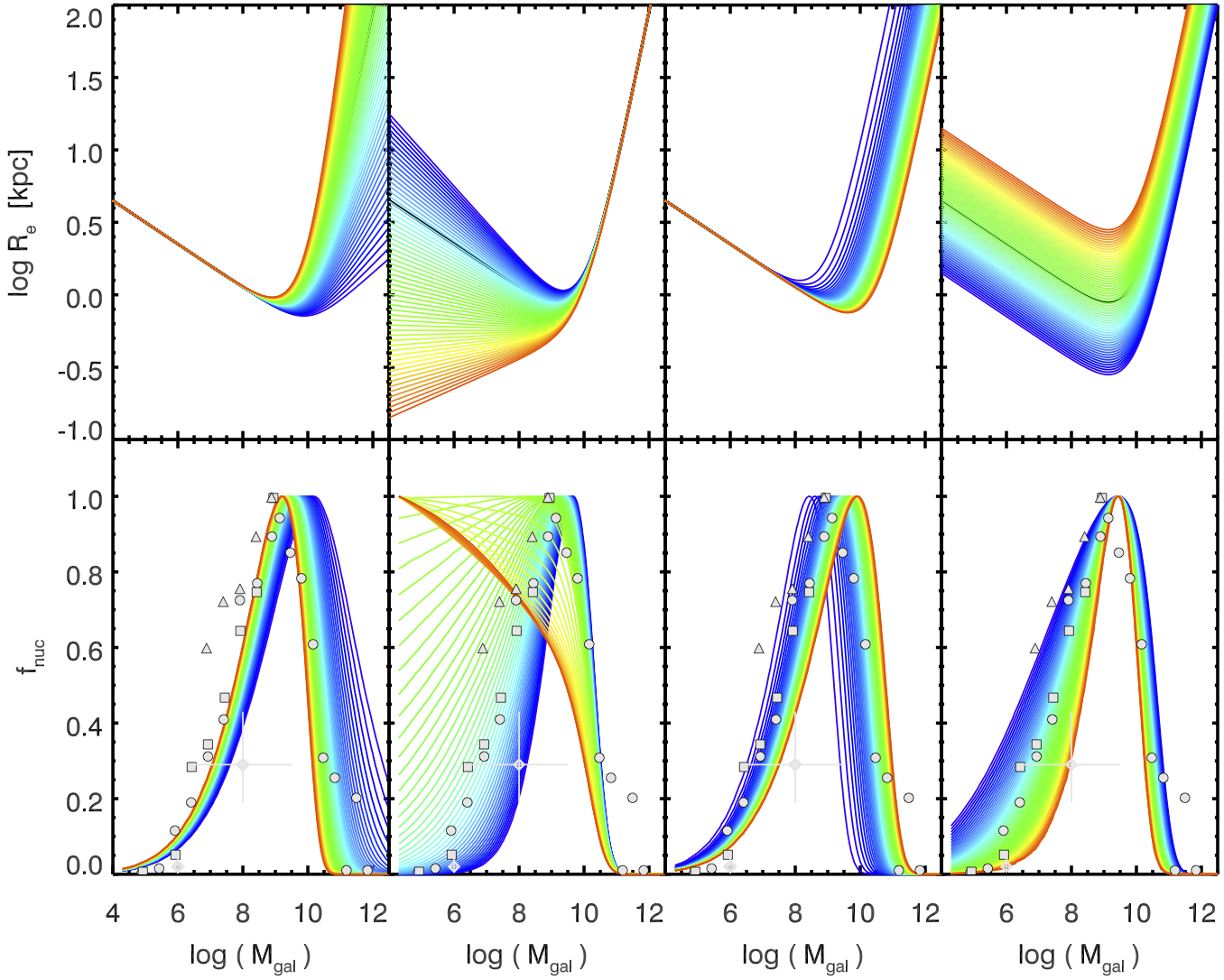}
\caption{Top row shows parameterized galaxy size-mass relations, with the corresponding changes to the model nucleation fractions in the bottom panels. }
\label{fig:f6var}
\end{figure*}

To quantify which aspects of the galaxy size-mass relation set the shape and peak location of the $f_{nuc}-M_{gal}$ relation, in Figure \ref{fig:f6var} we show a generic parameterized size-mass relation using the function presented in \cite{Leauthaud12} to describe the stellar-to-halo mass relation of galaxies.

In general, any region of the size-mass plane in which a variation produces larger galaxies, tends to have a corresponding decrease in the nucleation probability at that mass.

\subsection{Effects on the GC and NSC system mass}

\subsubsection{Dependence on fraction of clustered star formation}
\begin{figure}
\includegraphics[width=0.48\textwidth]{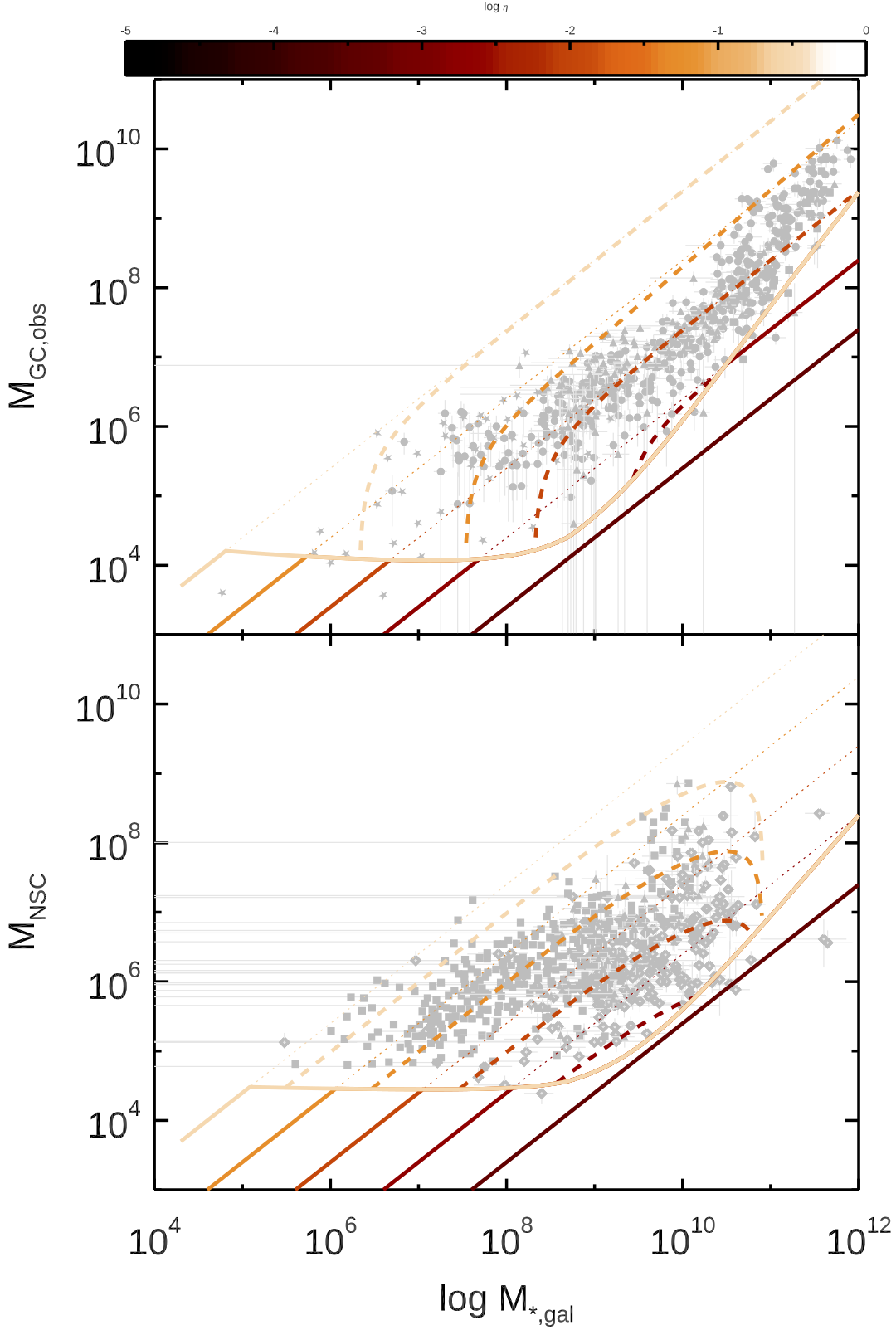}
\caption{Predicted mass in surviving GCs (\textit{top}) and mass of the host galaxy's NSC (\textit{bottom}) as a function of galaxy stellar mass.  Observational data in grey are from \protect{\citealt{Peng06,Spitler09,Harris91,Forbes18,RSJ19}}  Model lower limits, expectation values and upper limits are shown as the solid, dashed, and dotted lines respectively, and colour coded by the star cluster formation efficiency, $\eta$.}
\label{fig:obseta}
\end{figure}

Figure \ref{fig:obseta} shows the predicted NSC and GC system masses in our model for different values of the bound cluster formation efficiency, $\eta$.  The envelope and large intrinsic scatter of the GC and NSC masses is well encompassed by our models, with the lower limits in particular fitting the boundary of the most cluster deficient galaxies with no fine tuning.  In both the NSC mass and the GC system mass, the curved lower limit of our model is independent of the cluster formation efficiency until the point where that lower limit approaches the linear value of $\eta M_{*,gal}$.  In this regime a few of the lowest mass dwarf galaxies occupy regions where $\eta$ is uniquely dictating the galaxy's cluster population ($M_{*,gal} \lesssim 10^{7}$ and $M_{GC,obs} \lesssim 10^{4.5}$).  Such extreme systems can be compared to alternative observational estimates of the cluster formation efficiency.  Alternatively, their presence below the $\eta = 1.0$ lower limit may suggest that these low mass ($M_{GC} \lesssim  10^{4}$) star clusters are in the process of disruption and have undergone significant mass loss.

Reassuringly no galaxies show GC system or NSC masses above the \textit{upper} limits of $\eta = 1$ in either parameter space. The expectation values of the NSC mass reproduce the qualitative shape of the data distribution.  The turnover in the expectation value curves for the NSC mass is a function of the most massive GC mass formed in our model (Equation 10), here set to $M_{cl,max} = 10^{6.5}$.  The GC system masses are not expected to follow the slope of the expectation values from our model, as we do not consider the impact of accreted GCs and stellar mass from galaxy mergers (see \S 5.4).  Nevertheless, it is pleasing that the hard upper and lower limits of the model envelope are well respected by the current observations.

\subsubsection{Dependence on GC formation distance}
\begin{figure}
\includegraphics[width=0.48\textwidth]{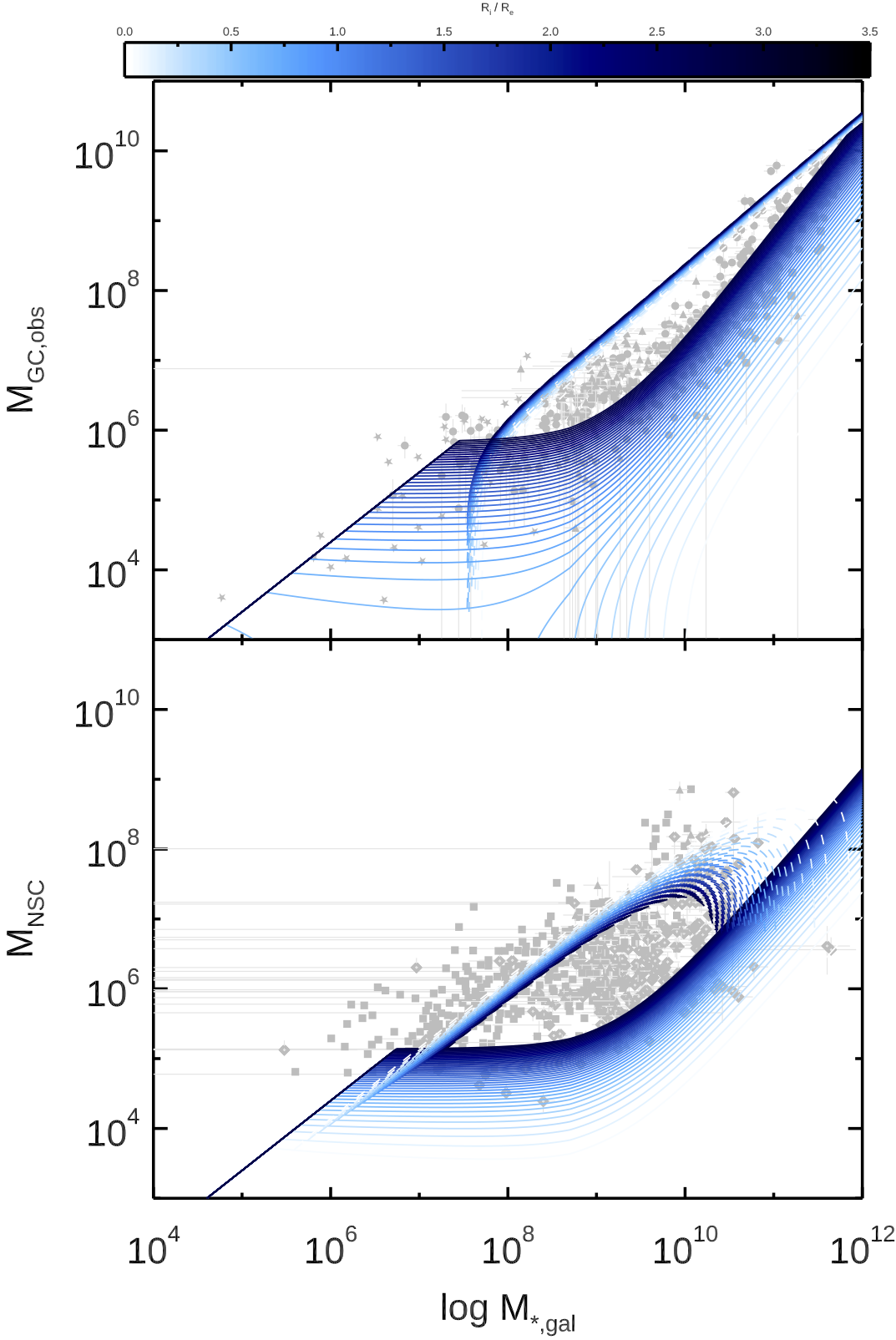}
\caption{Same as Figure \protect{\ref{fig:obseta}} but for variations in the initial GC formation distance, $R_{i}/R_{e}$ as a function of the galaxy effective radius.  Solid lines show lower limits, and dashed lines expectation values, predicted in our model. In models where the GCs formed proportionally closer to the host centre the average NSC mass is increased, while the minimum surviving GC mass decreases as more of the objects are deposited into the NSC.}
\label{fig:obsdi}
\end{figure}

In Figure \ref{fig:obsdi} we show how the NSC and GC system mass change as the initial formation distance of the in-spiraling GC's is changed.  The expectation value for the GC system mass is essentially independent of the formation distance.  However the lower limit to $M_{GC,obs}$ decreases as the formation distance is reduced.  We see in the lower panel that the expectation value of the {NSC} mass increases as the formation distance is reduced.  This could indicate that the smaller starting galactocentric distance aids GCs in successfully contributing to the NSC buildup (and simultaneously depleting the GC population).  However as this is primarily effecting the lowest mass GCs, the net effect on both expectation values is not substantial.  

An important point is that for fixed star cluster efficiency, the upper limit to the NSC and GC system masses is invariant to the GC formation distance.  This means in essence that a larger \textit{spread} in both scaling relations at fixed galaxy mass will occur when the GCs form initially closer to the galaxy centre.

\subsubsection{Dependence on GC mass function}
\begin{figure}
\includegraphics[width=0.48\textwidth]{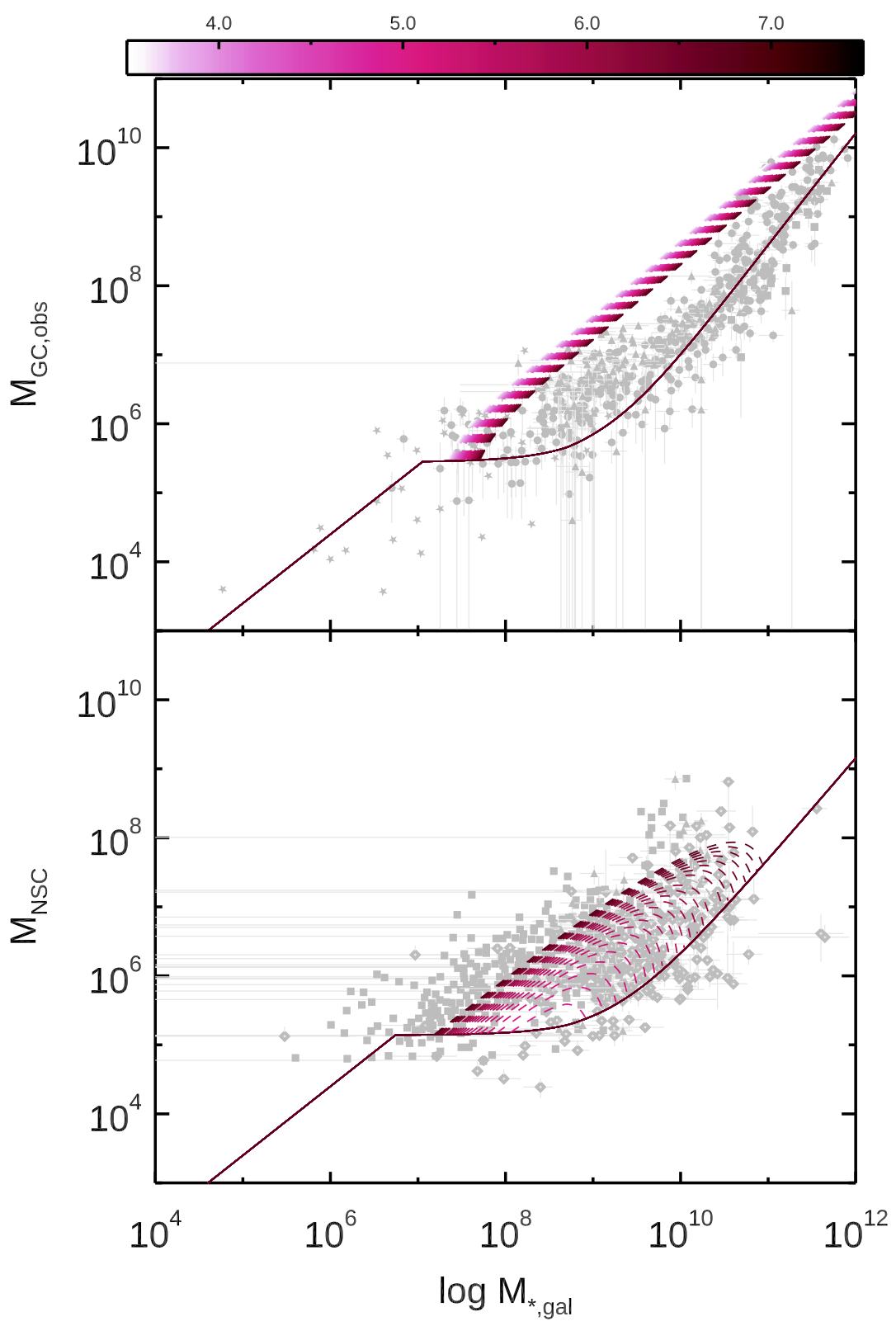}
\caption{Same as Figure \protect{\ref{fig:obseta}} but for variations in the most massive GC formed, $M_{GC,max}$.  Solid lines show lower limits, and dashed lines expectation values, predicted in our model. Galaxies which form higher mass clusters tend to grow larger NSCs, with a mild depletion in their expected total GC populations. The range of galaxy masses with significant NSCs increases significantly with this parameter.}
\label{fig:obsmmax}
\end{figure}

While the average GC mass impacts the width of the nucleation fraction function (\S 4.1.2), the NSC and GC system masses are much more strongly affected by the most massive GC formed by the galaxy (Equations 7, 10).  Figure \ref{fig:obsmmax} shows that increasing the cutoff mass of the GC mass function increases the expectation value for the total GC system mass and the NSC mass (at fixed galaxy mass).  Both the lower and upper limits are insensitive to the most massive cluster formed, only the expectation value changes (\S 3.3). 

Importantly, the maximum star cluster mass plays a strong role in setting the turnover galaxy mass where the NSC expectation values begin to decline.  If a galaxy forms star clusters only up to $M_{GC} \sim 10^{5}$, the models predict a narrower range of galaxy masses would host NSCs (the lowest dashed curve from $8 \lesssim {\rm log}M_{*,gal} \lesssim 9$).  Increasing the maximum cluster mass to $10^{7}$ results in a truncation to the distribution of host galaxy masses closer to the observed boundary at $M_{*,gal} \sim 10^{11}$.  

\subsubsection{Dependence on host galaxy structure}
\begin{figure}
\includegraphics[width=0.48\textwidth]{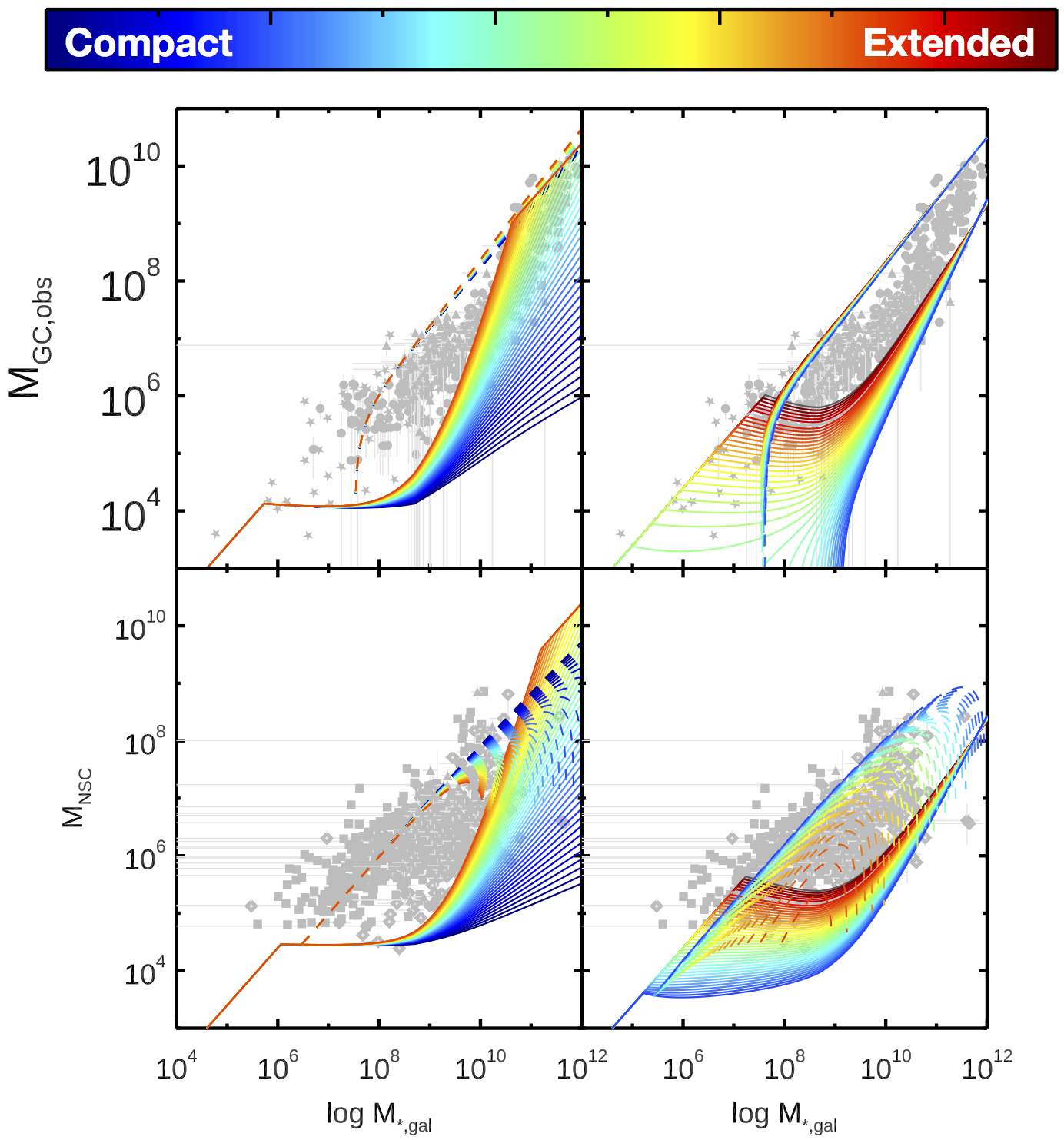}
\caption{Same as Figure \protect{\ref{fig:obseta}} but for variations in the high mass (\textit{left panels}) and low mass (\textit{right panels}) slopes of the galaxy size-mass relation.  Solid lines show lower limits, and dashed lines expectation values predicted in our model.}
\label{fig:obsrgal}
\end{figure}

Figure \ref{fig:obsrgal} shows the impact on the star cluster scaling relations if the slope of the low mass or high mass galaxy size-mass relation are altered.  In both cases we see that more extended galaxies have a higher threshold (lower limit) for the GC system \textit{and} NSC masses.  The expectation value of the GC system masses increases (albeit only slightly) for more extended galaxies.  

Conversely, the expectation value for the NSC mass is reduced in these systems, and the range of galaxies hosting substantial NSCs shows a similar behaviour when the galaxies are more extended, as when the maximum GC mass is truncated (bottom panel of Figure \ref{fig:obsmmax}).  This suggests that the typical GC mass which survives infall is higher in the extended galaxies (for fixed formation distance, $R_{i}/R_{e}$) due to the longer time/distance they must travel to contribute to the nucleus.  

Similar to Figure \ref{fig:obsdi}, the upper limit is insensitive to the galaxy size- mass relation.  This means that the \textit{spread} in NSC and GC system masses at fixed host mass will be reduced as galaxies become more extended. The spread in both scaling relations is maximal around $M_{*,gal}\sim 10^{9}$ and tapers on either side - consistent with this being the location of some of the most compact galaxies on average.  We refer the reader to \S C3 for a more complete discussion on the predicted spreads and expected \textit{relative} masses in GCs and NSCs that galaxies have in this model.

\subsection{Effects on the NSC to GC system mass ratio}

\subsubsection{Dependence on fraction of clustered star formation}
\begin{figure}
\includegraphics[width=0.48\textwidth]{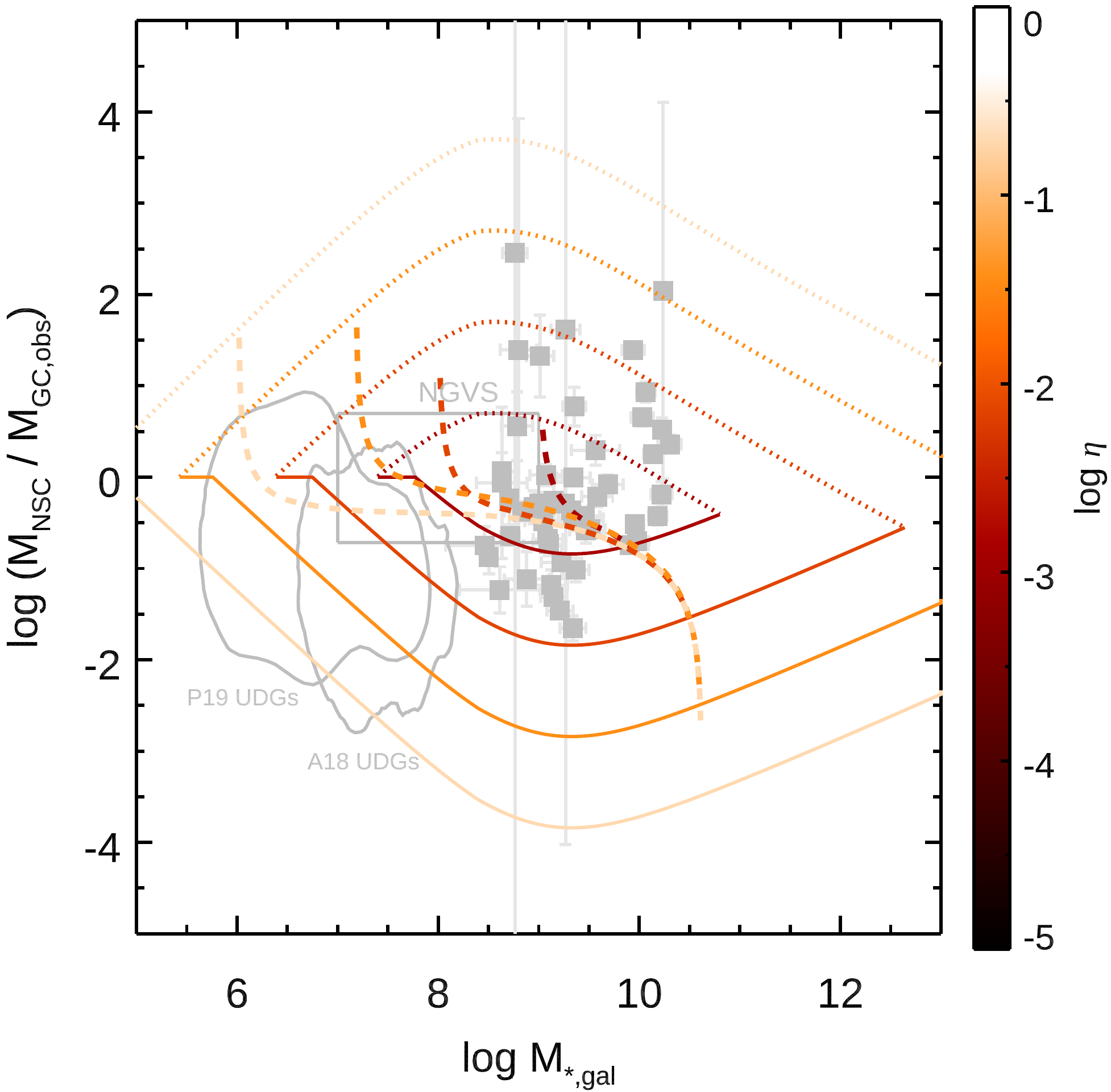}
\caption{Same as Figure \ref{fig:rat} but for variations in the fraction of star formation occuring in clusters, $\eta$.  Solid lines are lower limits, dotted lines upper limits and dashed lines expectation values.}
\label{fig:obsrateta}
\end{figure}

At lower values of $\eta$ Figure \ref{fig:obsrateta} shows that the expected range of galaxy masses which have both NSCs and GCs becomes restricted.  Smaller values of $\eta$ also result in a smaller intrinsic scatter in the ratio of $M_{NSC}/M_{GC,obs}$.

\subsubsection{Dependence on GC formation distance}
\begin{figure}
\includegraphics[width=0.48\textwidth]{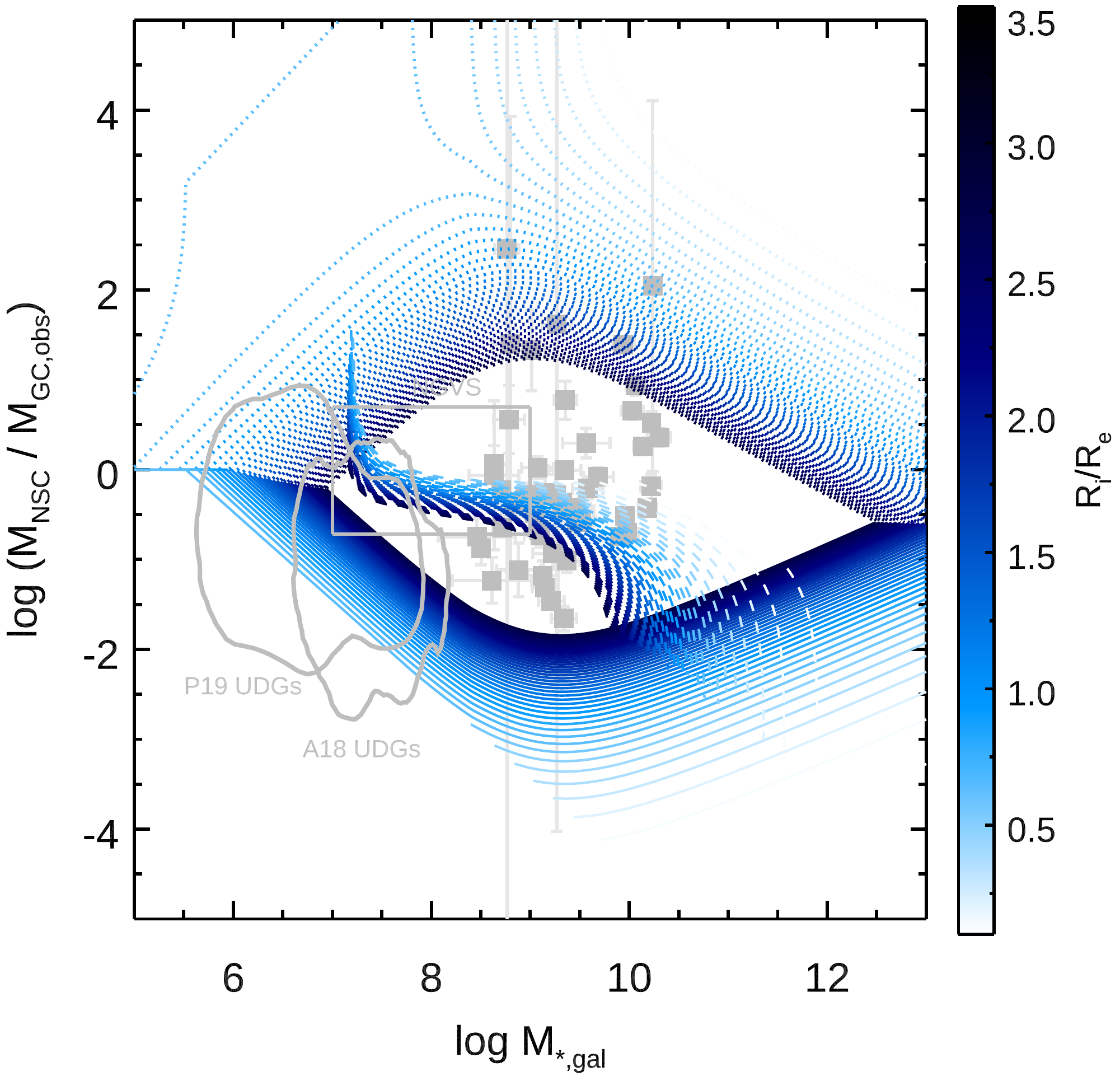}
\caption{Same as Figure \ref{fig:rat} but for variations in the GC initial formation distance, $R_{i}/R_{e}$.  Solid lines are lower limits, dotted lines upper limits and dashed lines expectation values.}
\label{fig:obsratdit}
\end{figure}

In Figure \ref{fig:obsratdit} we show the change to the ratio of NSC to GC system mass if the GC formation distance is varied.  As GCs form preferentially towards the inner regions of their host, our model predicts that the NSC mass becomes proportionally larger relative to the NSC system mass as the conditions for in-spiral are aided.  In addition the intrinsic spread in the ratio of $M_{NSC}/M_{GC,obs}$ is increased in the case of smaller $R_{i}/R_{e}$.

\subsubsection{Dependence on host galaxy structure}
\begin{figure}
\includegraphics[width=0.48\textwidth]{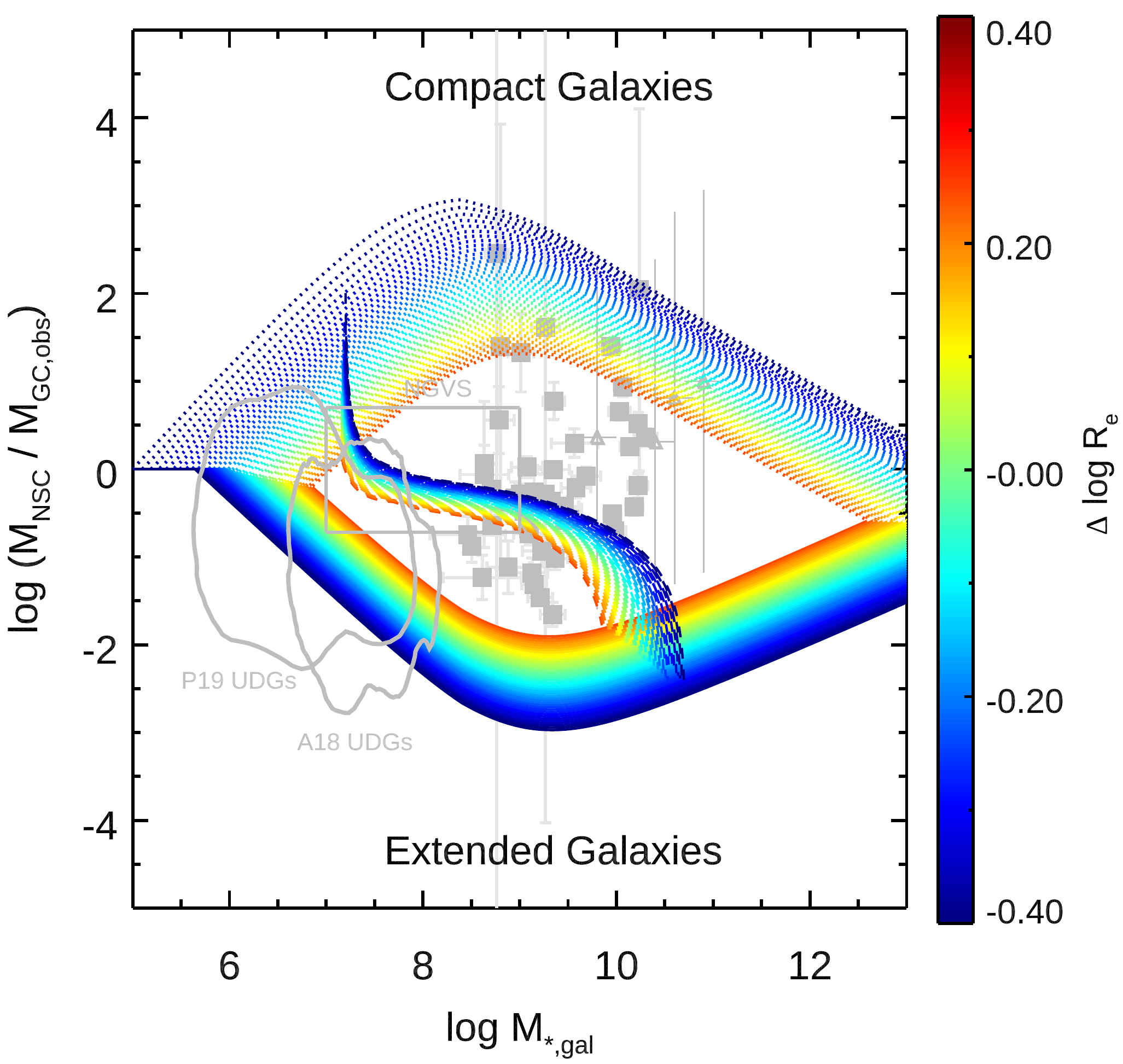}
\caption{Same as Figure \ref{fig:rat} but for variations in the host galaxy effective radius (at fixed stellar mass).  Solid lines are lower limits, dotted lines upper limits and dashed lines expectation values.}
\label{fig:obsratroff}
\end{figure}

Finally in Figure \ref{fig:obsratroff} we show variations in $M_{NSC}/M_{GC,obs}$ due to changes in galaxy size (at fixed stellar mass). Here we see that more compact galaxies have expectation values with larger values of $M_{NSC}/M_{GC,obs}$ relative to more extended systems.  The intrinsic scatter is mildly larger for more compact galaxies however.  The trend of the expectation values is in agreement with the observational data for UDG galaxies showing lower values of $M_{NSC}/M_{GC,obs}$ in this parameter space.  This small nuclei mass is consistent with at least two of the formation scenarios for UDGs - intrinsically large from birth due to occupying high spin DM halos \citep{Amorisco18}, or puffed up due to stellar feedback driven core creation \citep{diCintio17}.  In both cases the gas disk and potential in-spiral location of the GCs would be much larger in a physical sense than typical galaxies in the Local Volume, leading to less efficient in-spiral and lower nuclei masses.

\clearpage
\newpage
\newpage
\section{Dependence on \textit{ex-situ} star clusters and \textit{in-situ} NSC growth}\label{app:d}
Our model formally considers that GCs formed in the host galaxy and contribute to the mass growth of its central NSC.  Two factors which are known to influence both the total GC system mass and NSC mass are: (i) the accretion of stellar material and GCs by galaxy mergers, and (ii) the growth of the NSC through \textit{in-situ} star formation rather than in-spiral of massive star clusters.  

 In the case of galaxies which undergo significant number of mergers, the total mass of the galaxy will increase, and additional GCs will be added to the galaxy.  \cite{Elbadry19} showed that the evolution of the total GC system mass and galaxy mass proceed along the scaling relation defined by the two quantities.  \cite{Beasley18} also showed that the accreted fraction of stellar material brought in by mergers to the galaxy, scales monotonically and linearly proportional to the fraction of ex-situ GCs, e.g., $f_{acc,gal} \simeq f_{acc,GC} \equiv f_{acc}$.

Model predictions for the ratio of the NSC mass to the GC system mass can be augmented analytically to allow for some fraction of the galaxy's mass and GCs to be accreted ($f_{acc}$) and a fraction of the NSCs mass to be formed by \textit{in-situ} star formation ($f_{in,NSC}$).  In such a case limits for the fraction of the NSC mass which is formed through \textit{in-situ} star formation are:
 
\begin{equation}
    f_{in,NSC} \geq 1 - \left\lbrace\frac{\eta(1-f_{acc})^{2}M_{gal}M_{GC,obs}}{M_{NSC}M_{GC,lim}\left(1 + {\rm ln}\frac{M_{GC,lim}}{M_{diss}}\right)}\right\rbrace
\end{equation}

\begin{equation}
    f_{in,NSC} \leq 1 - \left\lbrace\frac{M_{GC,lim}M_{GC,obs}}{\eta M_{gal}M_{NSC}}\right\rbrace
\end{equation}

where $f_{acc}$ is the fraction of the galaxy's stellar mass and GCs brought in via galaxy mergers.  The expectation value for the NSC \textit{in-situ} fraction follows from Equation 15 and takes the form of:

\begin{equation*}
    f_{in,NSC} =  1 - \left\lbrace\frac{\eta M_{gal}M_{GC,obs}(1-f_{acc})^{2}
    \left[\frac{1 + {\rm ln}\frac{M_{cl,max}}{M_{GC,lim}}}{1 + {\rm ln}\frac{M_{GC,lim}}{M_{cl,min}}}\right]}{M_{NSC}\left(\eta M_{gal}(1-f_{acc}) \left(1 -\left[\frac{1 + {\rm ln}\frac{M_{cl,max}}{M_{GC,lim}}}{1 + {\rm ln}\frac{M_{GC,lim}}{M_{cl,min}}}\right]\right) - M_{diss}\left(1 + {\rm ln}\frac{M_{diss}}{M_{cl,min}}\right)\right)}\right\rbrace
\end{equation*}

\begin{figure*}
\includegraphics[width=0.48\textwidth]{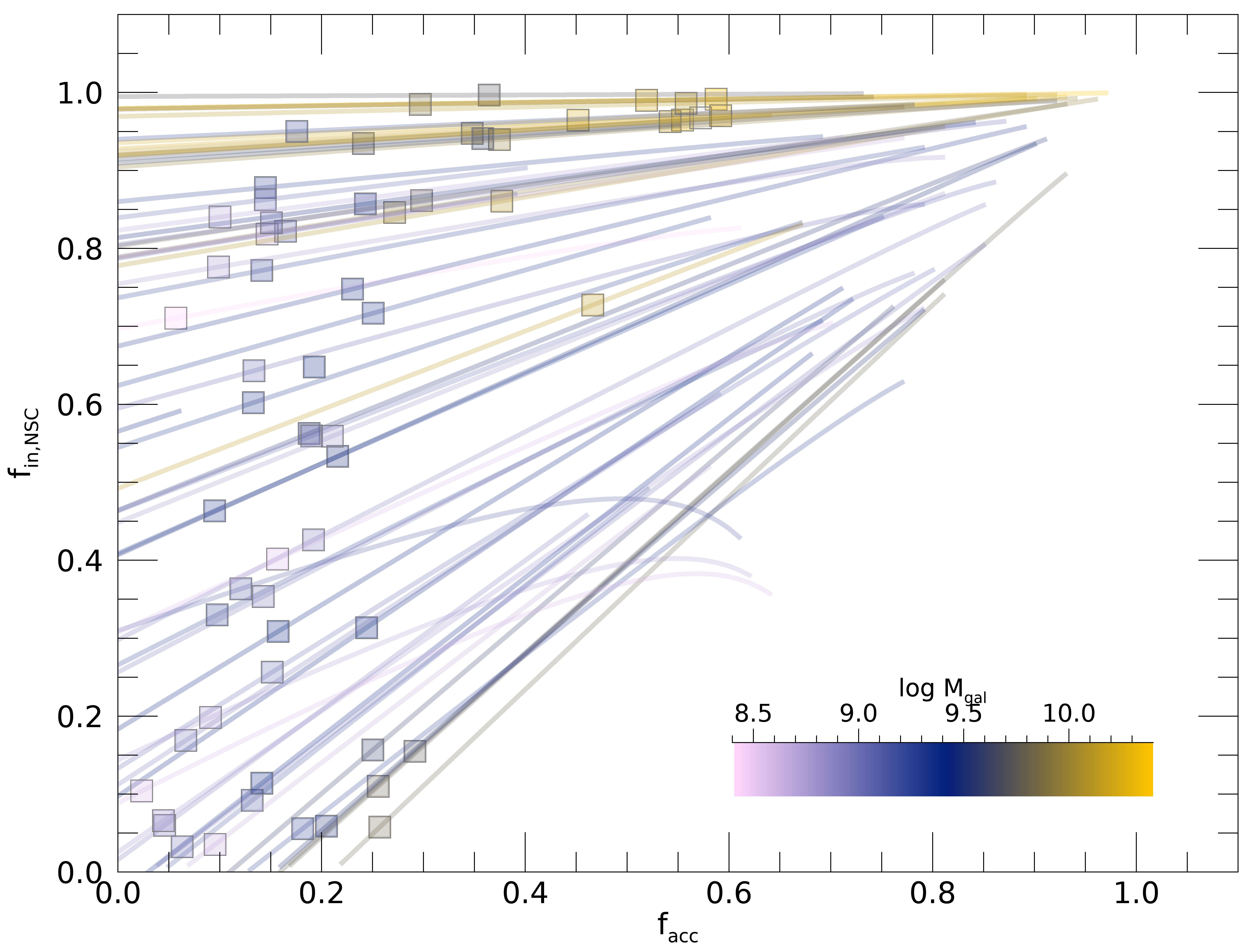}
\includegraphics[width=0.48\textwidth]{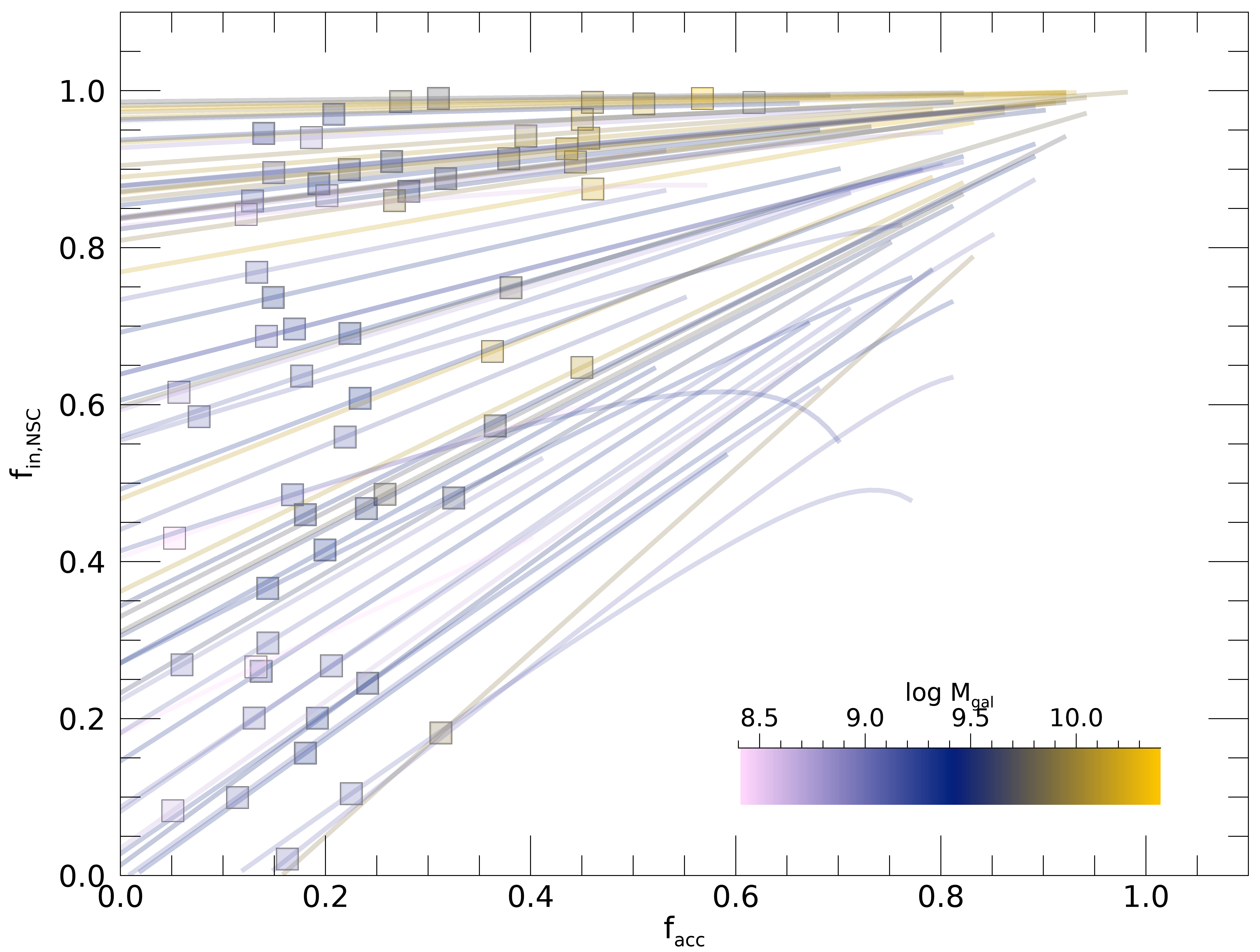}
\caption{Two realizations of mock galaxies with realistic values of GC system, NSC, limiting GC masses and sizes which are given random values of the NSC $in-situ$ fraction and GC accretion fraction (\textit{boxes}.  For each galaxy the coloured lines show the systematic uncertainty and model limits on $f_{in,NSC}$ as the accretion fraction changes ($f_{acc}$).}
\label{fig:inacc}
\end{figure*}

Figure \ref{fig:inacc} shows two realizations of how the the NSC \textit{in-situ} fraction responds to for a variable galaxy accretion fraction for a mock population of galaxies.  The galaxies were sampled normally with mean and standard deviation: ${\rm log}_{10}\eta = (-1.8, 0.5)$, ${\rm log}_{10}M_{GC,obs}/M_{gal} = (-1.9, 1.9)$, ${\rm log}_{10}M_{NSC}/M_{gal} = (-2.0, 1.7)$, ${\rm log}_{10}M_{cl,max} = (6.9, 1.6)$, ${\rm log}_{10}M_{cl,min} = (1.4, 0.3)$, ${\rm log}_{10}M_{GC,lim} = (6.0, 1.4)$ and ${\rm log}_{10}M_{GC,diss} = (4.6, 1.1)$.  These values are representative for galaxies in the Fornax and Virgo clusters (see Fahrion et al. in prep.)

The values for the galaxies are shown in this figure as the coloured squares while the co-dependence and analytic limits on $f_{in,NSC}$ as a function of the galaxy $f_{acc}$ are displayed as the coloured lines passing through each point.  Higher mass galaxies with high \textit{in-situ} NSC fraction would show a large possible range in the galaxy accretion history, but with a dependence which is rather shallow and limits the systematic change to $f_{in,NSC}$.  Some lower mass galaxies show large systematic variations in the $f_{in,NSC}$ were they to have large values of $f_{acc}$. However astrophysically it is not expected that lower-mass galaxies have had rich accretion histories to the same extent as higher mass galaxies.  


Suffice to say several galaxy mass dependent trends work together in this parameter space to maintain a more modest dependence between the NSC in-situ fraction and the galaxy accretion fraction than first appears.  Application of the model to estimate the NSC in-situ fraction in individual galaxies will be shown in an upcoming paper (Fahrion et al. in prep.)


\bsp	
\label{lastpage}
\end{document}